\documentclass[pageno]{jpaper}

\usepackage{multirow}
\usepackage{mathtools}
\usepackage{xspace}
\usepackage{ifthen}
\usepackage{calc}
\usepackage{pifont}
\usepackage{color}
\usepackage{fancyhdr}

\newcounter{hours}
\newcounter{minutes}

\newcommand{\hier}[0]{HIER\xspace}
\newcommand{\Bfive}{$B_{5\%}$\xspace}

\newboolean{timeofmake}
\setboolean{timeofmake}{false}

\newcommand{\ignore}[1]{}

\newcommand{\gabe}[1]{{\color{blue} \textbf{Gabe: #1}}\xspace}

\newcommand{\kevin}[1]{{\color{cyan} \textbf{Kevin: #1}}\xspace}
\newcommand{\lisa}[1]{{\color{green} \textbf{Lisa: #1}}\xspace}

\newcommand{\dontinclude}[1]{ }
\newcommand{\myfootnote}[1]{\footnote{\renewcommand{\baselinestretch}{1.0}\scriptsize #1}}

\newcommand{\putsec}[2]{\vspace{-0.05in}\section{#2}\label{sec:#1}\vspace{-0.075in}}
\newcommand{\putssec}[2]{\vspace{-0.05in}\subsection{#2}\label{ssec:#1}\vspace{-0.075in}}

\newcommand{\figput}[4][1.0\linewidth]{
\begin{figure}[t]
\begin{minipage}{\linewidth}
\footnotesize 
\begin{center}
\includegraphics[trim=#3, clip, width=#1]{plots/#2}
\end{center}
\caption{#4 \label{fig:#2}}
\end{minipage}
\end{figure}
}

\newcommand{\figputW}[4][1.0\linewidth]{
\begin{figure*}[t]
\begin{minipage}{\linewidth}
\footnotesize 
\begin{center}
\includegraphics[trim=#3, clip, width=#1]{plots/#2}
\end{center}
\vspace{-0.2in}
\caption{#4 \label{fig:#2}}
\end{minipage}
\end{figure*}
}

\newcommand{\figputT}[3]{
\begin{figure}[t]
\begin{minipage}{\linewidth}
\footnotesize 
\begin{center}
\includegraphics[trim=#2, clip, width=1.0\linewidth]{plots/#1}
\end{center}
\vspace{-0.2in}
\caption{#3 \label{fig:#1}}
\end{minipage}
\end{figure}
}

\newcommand{\figputTT}[3]{
\begin{figure}[t]
\begin{minipage}{\linewidth}
\footnotesize 
\begin{center}
\includegraphics[trim=#2, clip, width=1.0\linewidth]{plots/#1}
\end{center}
\vspace{-0.1in}
\caption{#3 \label{fig:#1}}
\end{minipage}
\end{figure}
}

\newcommand{\figref}[1]{Figure~\ref{fig:#1}}

\newcommand{\secref}[1]{Section~\ref{sec:#1}}

\usepackage[affil-sl]{authblk}
\makeatletter
\renewcommand\AB@affilsepx{ \quad \protect\Affilfont}
\makeatother

\newcommand{\squishlist}{
 \begin{list}{$\bullet$}
  { \setlength{\itemsep}{0pt}
     \setlength{\parsep}{1pt}
     \setlength{\topsep}{1pt}
     \setlength{\partopsep}{0pt}
     \setlength{\leftmargin}{1em}
     \setlength{\labelwidth}{1em}
     \setlength{\listparindent}{\parindent}
     \setlength{\labelsep}{0.5em} } }

\newcommand{\squishend}{
  \end{list}  }

\newcommand{\titleShort}[0]{CRUNCH\xspace}
\newcommand{\titleLong}[0]{Cache Resizing Using Native Consistent Hashing\xspace}

\author[$\dagger$]{Kevin K. Chang}
\author[$\star$]{Gabriel H. Loh}
\author[$\ddagger$]{Mithuna Thottethodi}
\author[$\star$]{Yasuko Eckert}
\author[$\star$]{Mike O'Connor}
\author[$\star$]{\authorcr Srilatha Manne}
\author[$\star$]{Lisa Hsu}
\author[$\dagger$]{Lavanya Subramanian}
\author[$\dagger$]{Onur Mutlu}
\affil[$\dagger$]{Carnegie Mellon University}
\affil[$\star$]{AMD Research}
\affil[$\ddagger$]{Purdue University}

\usepackage[T1]{fontenc}
\usepackage{libertine}

\usepackage{pifont}

\begin{document}

\title{Enabling Efficient Dynamic Resizing of Large DRAM Caches via\\ A Hardware Consistent Hashing Mechanism}

\date{}
\maketitle

\thispagestyle{plain}
\pagestyle{plain}

\begin{abstract}

Die-stacked DRAM has been proposed for use as a large,
high-bandwidth, last-level cache with hundreds or thousands of
megabytes of capacity.  Not all workloads (or phases) can
productively utilize this much cache space, however.
Unfortunately, the unused (or under-used) cache continues to
consume power due to leakage in the peripheral circuitry and
periodic DRAM refresh. Dynamically adjusting the available DRAM
cache capacity could largely eliminate this energy overhead.
However, the current proposed DRAM cache organization introduces
new challenges for dynamic cache resizing. The
organization differs from a conventional SRAM cache organization
because it places entire cache sets and their tags within a single
bank to reduce on-chip area and power overhead. Hence, resizing a
DRAM cache requires remapping sets from the powered-down banks to
active banks. 

In this paper, we propose \titleShort(\titleLong), a hardware data
remapping scheme inspired by consistent hashing, an algorithm
originally proposed to uniformly and dynamically distribute
Internet traffic across a changing population of web servers.
\titleShort provides a load-balanced remapping of data from the
powered-down banks {\em alone} to the active banks, without
requiring sets from {\em all} banks to be remapped, unlike naive
schemes to achieve load balancing. \titleShort remaps only
sets from the powered-down banks, so it achieves this load balancing
with low bank power-up/down transition latencies. \titleShort's
combination of good load balancing {\em and} low transition
latencies provides a substrate to enable efficient DRAM cache
resizing.  

\end{abstract}

\putsec{intro}{Introduction}
Die-stacking technologies are rapidly
maturing~\cite{kim-isscc2011,pawlowski-hotchips2011}. One likely
near-term use is to stack a processor with a
large, high-bandwidth, in-package DRAM
cache~\cite{black-micro2006,jiang-hpca2010,loh-micro2011}.
Projections indicate that the size of the DRAM cache may be
hundreds of megabytes or more.  Many workloads (or workload
phases) may not productively utilize such large caches, however.
This leads to wasted energy consumption because of significant
DRAM background power (from leakage and peripheral circuitry).
Recent studies have shown that a system's DRAM is consuming an
increasing fraction of the overall system
power~\cite{hoelzle-book2009,meisner-asplos2009}. While a
die-stacked DRAM's overall capacity will be less than that of main
memory (and therefore not likely to consume nearly as much power
as off-chip DRAM), in the present age of power-constrained
designs, any power consumed by the stacked DRAM is power that
cannot be consumed by the CPU cores~\cite{gunther-itj2010}.

Thus, there is opportunity to reduce the energy consumption of the
die-stacked DRAM cache while maintaining the performance
advantages of having a large-capacity caching structure. When the
DRAM cache is under-utilized, it should reduce its
active portions to match the capacity needs of
the current workload. This can be achieved by turning
off some banks of the DRAM cache. While turning off ways/banks is a
well-studied problem in the context of SRAM
caches~\cite{albonesi-micro99,Naveh:2006, powell-islped2000,
zhang-isca2003}, the organization of the DRAM cache
poses new challenges. Current proposals for DRAM cache
organizations~\cite{dong-sc2010,loh-micro2011,timber} call for
entire sets and their tags to reside within a single bank in order
to reduce the number of row activations per access. Hence, a DRAM
cache resizing mechanism would need to (1) consider not only which
banks to turn off (well-explored problem in SRAM caches) but also
(2) address the new challenge of remapping sets from the
\emph{powered-down} banks into active banks (or suffer 100\% miss
rates for those addresses) and migrating the dirty data from the
powered-down banks to either the active banks or to the off-chip
DRAM. 

Naive data remapping when resizing a DRAM cache can
be very costly. One possible way to do
this is to remap all of the data from a powered-down bank into
another active bank. This remapping scheme could make
the active bank a hotspot, increasing DRAM cache access latencies.
Another possibility is to completely remap data from all
banks to the remaining active banks with a
modulo-$k$ operation ($k$=number of active banks after a bank
shut-down), providing a uniform distribution of data among the
active banks without creating hotspots. However, completely
remapping data across all banks every time the cache size changes can be
very costly, as this requires migrating a large amount of dirty
data. 

To address the challenge of remapping data 
in a load-balanced manner with
low transition overhead, we take inspiration from the
{\em consistent hashing} algorithm originally proposed to
uniformly and dynamically distribute load across a large number of
web servers~\cite{karger-stoc97}. Cloud providers may have servers
go down at any point in time (e.g., server crashes,
scheduled maintenance). Similarly, new servers may come
online (e.g., machines rebooted, adding new
machines to the server pool). The frequency of
machines coming and going make it infeasible to perform a complete
re-indexing each time the pool of available servers changes.

Consistent hashing provides an elegant scheme where a
machine failure results in re-indexing {\em only the subset of
elements that mapped to that machine}. That subset is
redistributed among the surviving machines in a load-balanced
manner; at the same time, any elements already mapped to the
remaining machines remain where they are. Turning off DRAM cache
banks is analogous to servers going down (and turning on
banks is like servers coming online).  In this work, we
present a hardware remapping scheme inspired by consistent
hashing, called \emph{\titleLong(\titleShort)}, to maintain
load balancing among the remaining active banks while
providing efficient transitions. 

Our paper makes the following contributions:

\begin{itemize}
\item To our knowledge, this is the first paper to observe and
address the challenges of remapping data from powered-down banks 
in DRAM caches in a load-balanced manner, thereby
\textit{enabling dynamic DRAM cache resizing}.
\item We propose a low-overhead mechanism, CRUNCH, to remap data
from powered-down banks. 
Our mechanism, inspired
by consistent hashing, incurs low latency/power overhead, while
achieving good load balancing.
\item We compare our mechanism, CRUNCH, to two data mapping
schemes, one optimized for low transition overhead and the other
optimized for load balancing. CRUNCH achieves the best of both
worlds, resulting in both good steady-state performance and
low transition costs.
\end{itemize}

\putsec{background}{Background and Motivation}

In this section, we provide a brief characterization of
application working set sizes, demonstrating the need
for DRAM cache resizing. We then discuss
DRAM cache organizations and two simple DRAM cache bank
remapping schemes to illustrate the challenges in powering down DRAM
cache banks.

\begin{figure}
\begin{center}
\begin{tabular}{c}
\hspace{-0.2in} \includegraphics[trim=0.7in 1.1in 1.6in 4.0in, clip, width=1.0\linewidth]{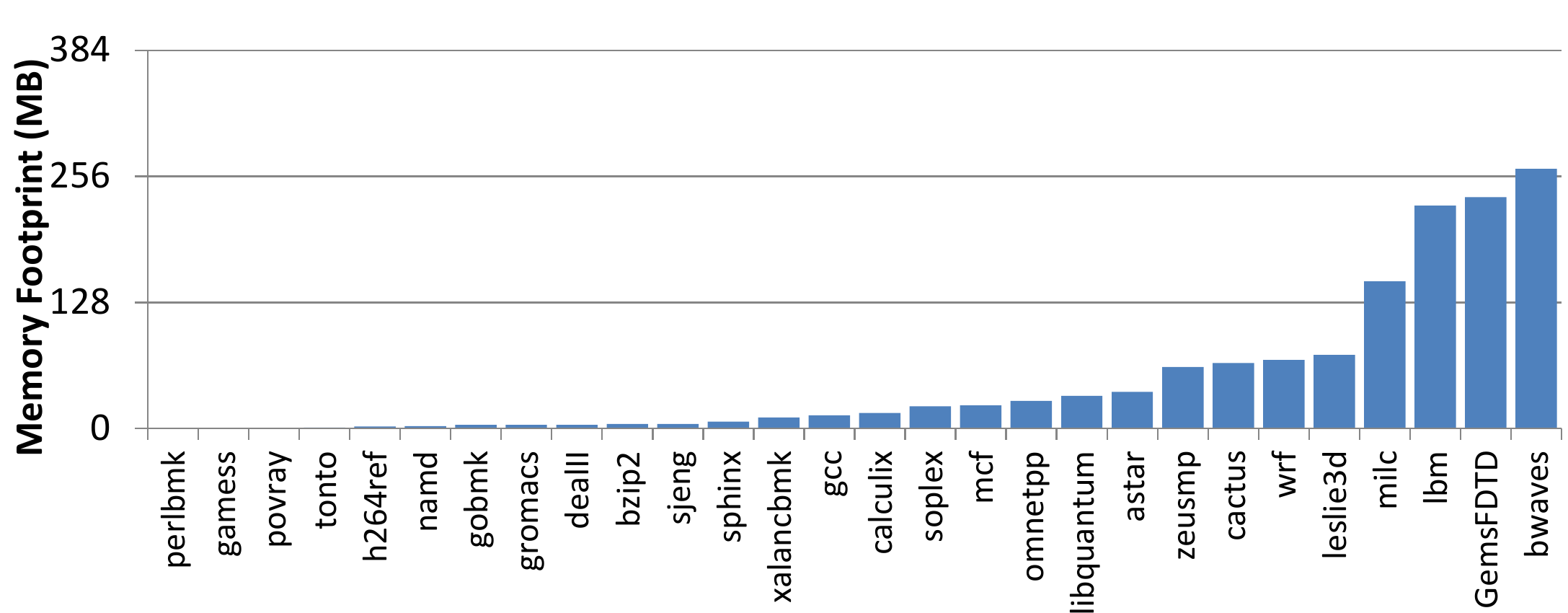} \\ {\footnotesize\sf\bfseries (a)} \\[-0.1in]
\hspace{-0.2in}\includegraphics[trim=1.0in 1.1in 1.2in 4.5in, clip, width=1.0\linewidth]{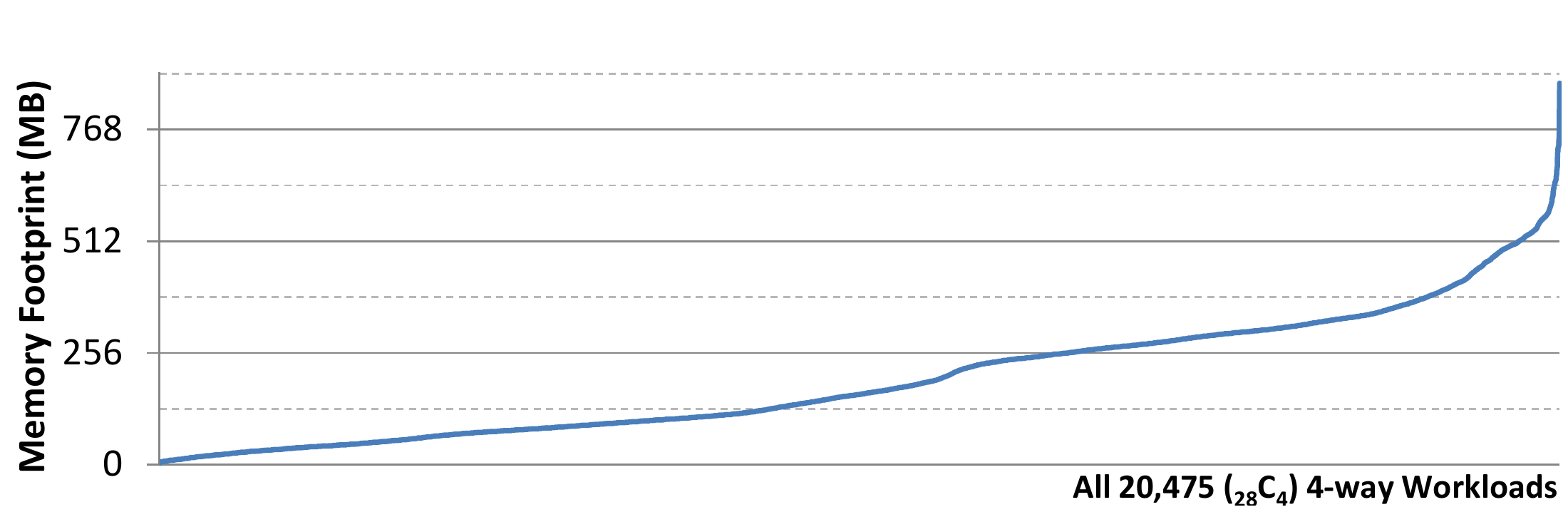} \\[-0.1in] {\footnotesize\sf\bfseries (b)}
\end{tabular}
\caption{Memory footprints for (a) each of the SPEC CPU2006 applications, and (b)
the aggregate footprint for each of the $_{28}C_{4}$ possible four-application
multi-programmed workloads.\label{fig:WSS}}
\end{center}
\end{figure}

\putssec{WSS}{Memory Footprint vs.~Cache Capacity}
Variations in working set sizes across applications, in
addition to variations in cache activity across time (program
phases), make it such that the full capacity of a DRAM cache is not always needed.
\figref{WSS}(a) shows the observed memory footprints for SPEC CPU2006
applications.\myfootnote{Simulation methodology is described in
\secref{meth}.} A few applications exhibit memory usage with large
footprints that would benefit from large DRAM caches. However, in many
other cases, the working sets would fit comfortably in a 128MB or
smaller DRAM cache. 
%
Modern machines have multiple cores capable of simultaneously
running multiple applications, and an effective DRAM cache should
be able to accommodate the {\em aggregate} memory demands of
multiple applications. \figref{WSS}(b) shows the total working set
sizes for all possible 4-core multi-programmed combinations of the
SPEC2006 workloads. With multiple simultaneously-running
applications, there is still a wide variance in memory demands,
with many combinations unable to fully utilize a large DRAM cache.
For a 128MB DRAM cache, 44\% of all 4-way workload combinations
would use less than the full cache capacity; for a 256MB DRAM
cache, 65\% of the combinations do not use the full capacity.
Therefore, although the generous capacity of die-stacked DRAM
caches is beneficial for some workloads/applications, there
clearly exist many scenarios where such a large cache is an
overkill. 

\ignore{
For a 128MB DRAM cache, more than X\% of all the 4-way workload
combinations use only Y\% of the DRAM cache capacity. For a 256MB
DRAM cache, more than X1\% of all the combinations use only Y1\%
of the DRAM cache capacity. Therefore, although the generous
capacity of die-stacked DRAM caches is beneficial for some
workloads/applications, there clearly exist many scenarios where
such a large cache is an overkill. These scenarios present ample
opportunity to turn off part of the DRAM cache. However, DRAM cache resizing is
not as straightforward as it might seem. As we will discuss in the
next two sub-sections, the way DRAM caches are organized presents
new challenges in dynamic resizing.
}

\ignore{
For a 128MB DRAM cache, 44\% of all 4-way workload combinations
would use less than the full cache capacity; for a 256MB DRAM
cache, 65\% of the combinations do not use the full capacity. This
argues that {\em at times} the generous capacity of die-stacked
DRAM would be beneficial, but it also highlights that there exists
many scenarios where such a large cache is an overkill.
}

\putssec{cache-basics}{DRAM Cache Organizations}


\figputT{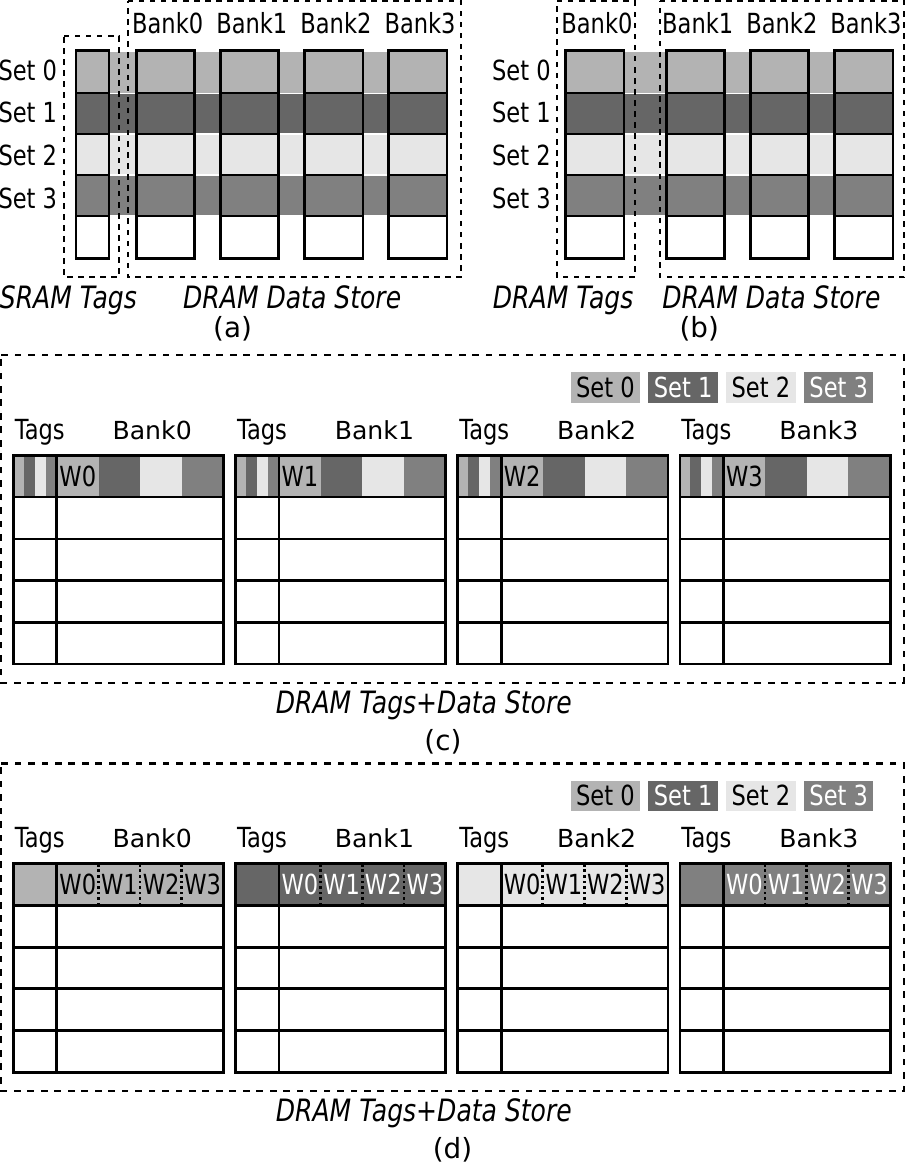}{0in 0in 0in 0in}{Different tags and data organizations
for DRAM caches: (a) tags in SRAM and data in DRAM,
(b) one dedicated DRAM bank for tags and the remaining banks for data,
(c) tags in DRAM with sets distributed across banks, and
(d) tags in DRAM with sets in a single bank. Each bank in (c) and (d) is the
same as a bank in (a) and (b). They may appear different because we are showing
a more detailed cache set-layout in (c) and (d).}

\noindent\textbf{SRAM Tag Store and Multi-Banked Cache Sets:}
One simple approach to organize tags and data in a DRAM cache is
to follow the conventional layout for a multi-banked SRAM cache as
shown in \figref{cache-4configs}(a). This organization has a
dedicated on-chip storage (i.e., SRAM) for tags\myfootnote{A tag
store can be single- or multi-banked.} and it splits the data into
multiple banks, thus each cache set is distributed across multiple
banks for a set-associative cache.  However, using an on-chip tag
store incurs tens of MB of overhead due to
the large projected capacities of DRAM caches~\cite{jiang-hpca2010,loh-micro2011}. 
As a result, a
conventional SRAM-style organization is impractical for
implementing a DRAM cache.

\noindent\textbf{Tags in one DRAM bank and Multi-Banked Cache
Sets:} To eliminate a large SRAM tag store, tags can be stored in
one/more DRAM banks with data distributed across the remaining
banks as shown in \figref{cache-4configs}(b). There are two main
disadvantages with this organization. First, performance will
suffer as the tag banks becomes a severe bottleneck, because
\emph{all} requests must access these banks for tag-lookup.
Furthermore, DRAM cannot multi-port or pipeline accesses like SRAM
does, because only one row may be opened at a time. Second, the
power consumption increases linearly with the number of tags banks because each
access now requires activating multiple banks for tags and data.

\noindent\textbf{Tags in DRAM and Multi-Banked Cache Sets:}
Another alternative to eliminate a large SRAM tag store is to
store tags in the DRAM in the same row as their data.
\figref{cache-4configs}(c) shows a DRAM cache using an SRAM-like
layout with tags and data distributed across all banks. Using this
organization has two main disadvantages. First, every cache line
lookup requires accessing \emph{all} banks in parallel, which
significantly increases the dynamic power consumption. This is
because every lookup requires sending one row activation to each
bank. Second, the opportunity to serve multiple cache line lookups
in parallel from different banks (i.e., bank-level
parallelism~\cite{tcm}) reduces because \emph{all} banks are
required to serve one lookup.

\noindent\textbf{Tags in DRAM and Single-Banked Cache Sets:}
To reduce the number of costly row activations required to perform an access,
Loh and Hill~\cite{loh-micro2011} proposed the DRAM cache organization shown in
\figref{cache-4configs}(d). This organization packs data and tags of a set
together in the same physical DRAM row/page within a bank.
\figref{isca/cache-config-power-crop.pdf} shows the power consumption of DRAM
caches with this
organization and the multi-banked cache sets organization (shown in
\figref{cache-4configs}(c)) across multiple workloads\myfootnote{Simulation methodology is described in
\secref{meth}.}. On average, the multi-banked design consumes 25.9\% more power
than the single-banked design proposed by Loh and
Hill~\cite{loh-micro2011}\myfootnote{Although workload 10 has very low memory intensity
(<2 L3-MPKI), it still increases the power by 9.3\% when using the multi-banked
cache sets design.}.
Therefore, unless stated otherwise, we assume a DRAM cache organization proposed
by Loh and Hill for the remainder of this paper.

\figputT{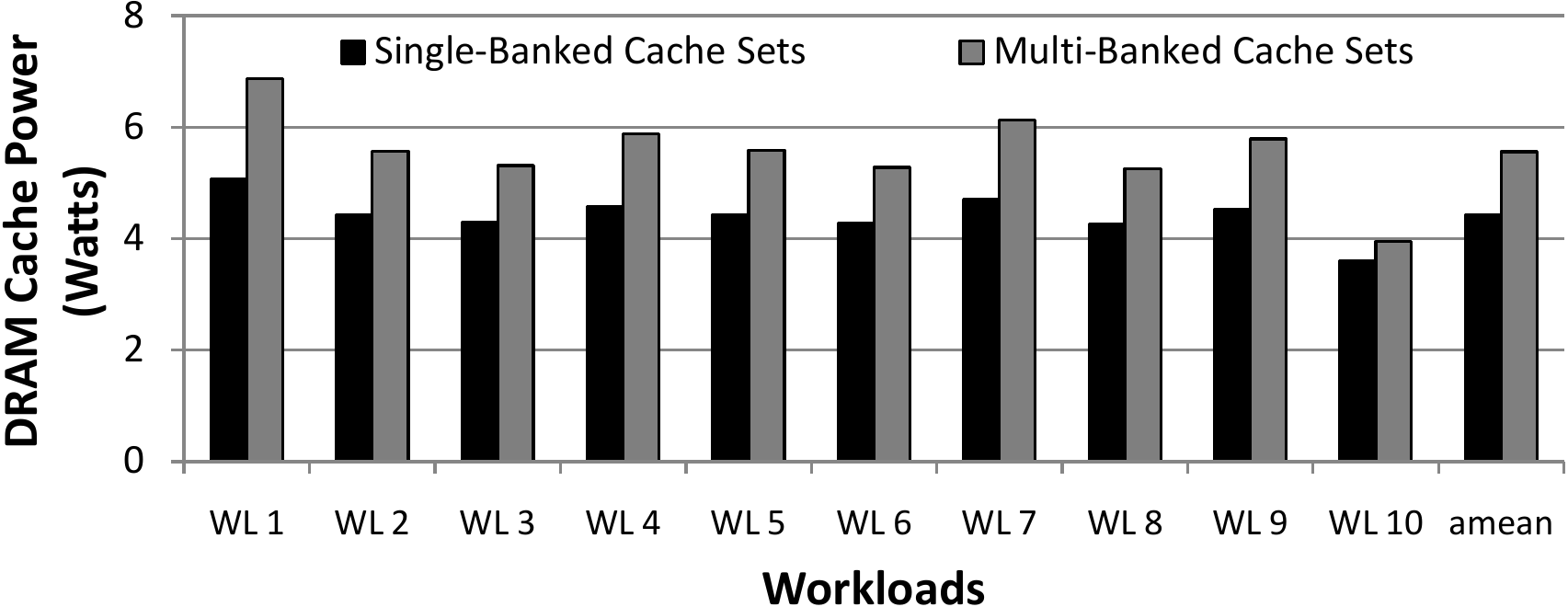}{0in 0in 0in 0in}{DRAM cache power
consumption comparison between the multi-banked cache sets design (shown in
\figref{cache-4configs}(c)) and the single-banked cache sets design (shown in
\figref{cache-4configs}(d)).}

\putssec{set-indexing}{DRAM Cache Resizing Challenges}

Prior works have studied resizing caches based on shutting down ways or banks~\cite{albonesi-micro99,Naveh:2006,powell-islped2000,
zhang-isca2003}. However, there are two main
challenges with dynamically resizing a DRAM cache.
First, resizing a cache with the Loh and Hill cache organization
by powering down banks introduces a
set-indexing issue because turning off a bank takes away all sets from
that powered-down bank. To avoid 100\% miss rates for
those requests that access cache sets in a powered-down bank, we need to
\emph{remap} these requests' addresses to active banks. 
Although using organizations that have sets distributed
across multiple banks as shown in \figref{cache-4configs}(a), (b), and (c) does not have
set-indexing issues for dynamically resized caches, these organizations are not
practical for large DRAM caches as discussed in the previous sub-section.
The second challenge is handling dirty data in powered-down banks, which
must either be written to memory or migrated to other banks\myfootnote{Clean
data can be safely dropped, which is what we assume in this paper.}. More data
migration incurs more overhead. We now discuss two simple remapping schemes
to illustrate these challenges. 

%
%
%


\subsection{Two Simple Bank Remapping Schemes}
\label{sec:naive}



\figputW{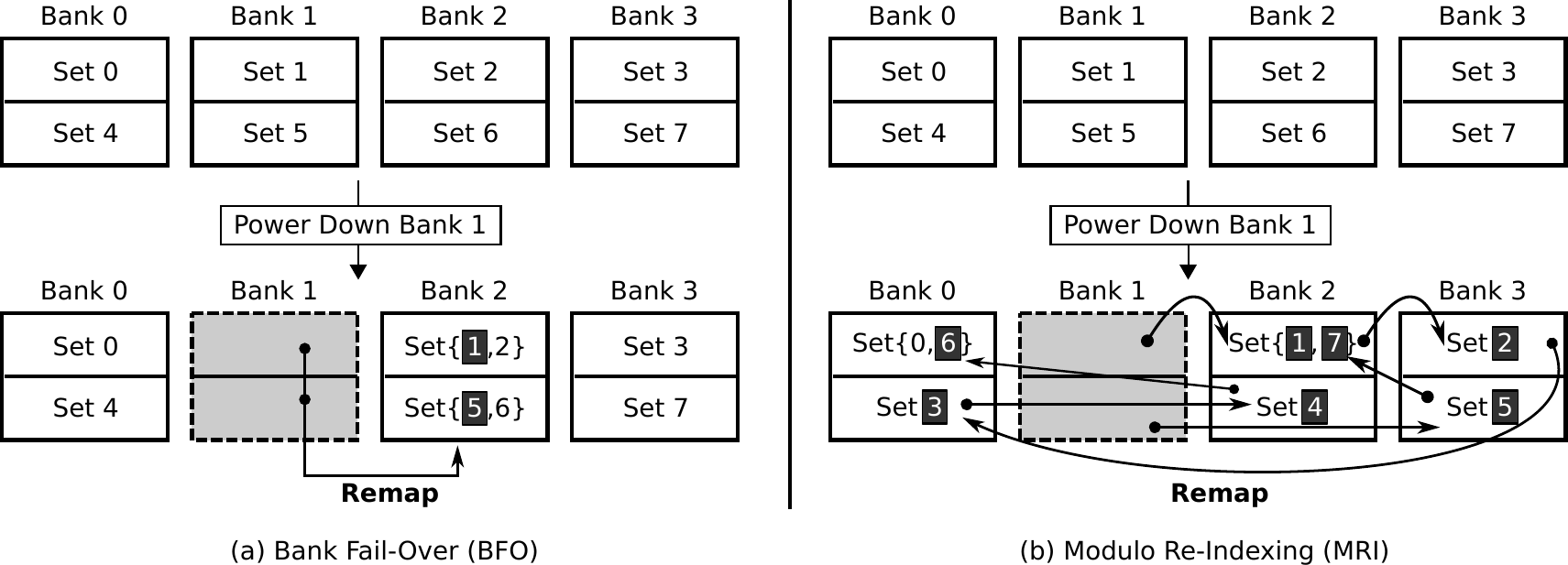}{0in 0in 0in 0in}{(a) An example scenario for Bank Fail-Over
where cache lines mapped to a powered-down bank are remapped to the next
sequential active bank. Dark blocks (white text) indicate sets that are remapped from
one bank to another. (b) An example scenario for Modulo Re-Indexing where the
address is rehashed (mod~3) and then remapped to a new bank. }


\noindent\textbf{Bank Fail-Over:}
The first scheme is \mbox{{\em Bank Fail-Over}} (BFO), where
cache lines mapped to bank $i$ simply get remapped, or fail over,
to the next available consecutive bank (with wrap-around).
\figref{mapping}(a) shows an example DRAM cache with four
total banks with one bank turned off. Cache lines that map to
bank~1 (which is off) are failed over to bank~2 (displaced sets are shown as
dark blocks in the figure). Such fail-overs
are easily computed by using the bank-selection bits from the
address in conjunction with a vector that indicates which banks
are active. In addition, because BFO allows addresses from
multiple banks to be remapped to the same bank, the cache tags
must be widened to include the bank-selection bits. 

The advantage of BFO is that when a bank is turned
off, \emph{only cache lines in the powered-down bank are
impacted}. In particular, BFO moves only dirty cache lines
to the fail-over bank. In the worst case, a 100\%
dirty bank requires migration or writeback of all of its
cache lines, but no other banks will be affected (apart from having
to accept the incoming migrants).

The disadvantage of BFO is that after bank shut-down,
there exists the potential for
\emph{unbalanced load distribution} across banks. For example
in \figref{mapping}(a), after bank~1 goes down, bank~2 has
twice as much capacity pressure as the other banks 
from having to
accommodate bank~1's cache lines as well as those originally
mapped to bank~2. This can lead to a sharp increase in conflict
misses for the affected banks.

\noindent\textbf{Modulo Re-Indexing:}
The second scheme, {\em Modulo Re-Indexing} (MRI), redistributes {\em all}
cache lines evenly across the remaining banks. In the top of
\figref{mapping}(b), all four banks are on, and the bank index for a given line
is computed by taking some number of bits from its physical address and
performing a modulo-4 operation. With MRI, when there are only $k$ banks
enabled, we compute the bank
index by performing a modulo-$k$ operation ($k$=3 in this example) and then
shifting the computed bank index to map to the actual enabled bank's index.
Other than the modulo computation hardware for re-indexing, MRI's hardware
requirements are similar to that of BFO (wider tags and bank selection from
among enabled banks).

The advantage of MRI is that cache lines are uniformly redistributed across all
banks, so no bank will be significantly more prone to being a hotspot
than any other. 
The disadvantage is that $k$ (number of active banks)
changes as banks are powered-down or -up, thus the majority of cache lines
will be remapped to new banks {\em on every transition}. 
This global reshuffling of cache contents, while good for load balancing,
severely increases the latency for powering down a bank, as nearly all dirty
cache lines migrate to their newly assigned
banks\myfootnote{Alternatively, they can be written back to main memory, and
then on the next miss reinstalled into the newly assigned bank.}. Thus, MRI's
power-down transition latency is proportional to the dirty data contained across \emph{all}
banks. 

The descriptions above on power-down transition latency also apply to bank
\emph{power-up} transition latency. With BFO, dirty cache lines in the fail-over
bank must be ``repatriated'' back to the bank that is being powered up. On the
other hand, with MRI, cache lines from all of the banks must be re-shuffled using
the updated modulo-($k$+1) mapping.

\noindent\textbf{Design Objective:}
Each of BFO and MRI have their strengths and
weaknesses, which are summarized in Table~\ref{summary}. BFO has bank
power-down/up latency proportional to only the number of \emph{powered-down}
banks, but suffers from poor load-balancing. MRI achieves uniform
load balancing, but suffers from poor transition latencies
proportional to the {\em total} number of active banks. In this work, we propose
\emph{\titleLong(\titleShort)}, a new DRAM cache remapping scheme, to
simultaneously achieve \emph{both} fast transition latency and load-balanced
cache line distribution.

\begin{table}
\centering
\begin{footnotesize}
\begin{tabular}{ccc}
\toprule
{\sf\bfseries Scheme} & {\sf\bfseries Fast Transition Latency} & {\sf\bfseries Load Balancing} \\
\cmidrule(rl){1-3}
Bank Fail-Over  & Yes & No  \\
\cmidrule(rl){1-3}
Modulo Re-Indexing & No  & Yes \\
\cmidrule(rl){1-3}
{\bf \titleShort} & {\bf Yes} & {\bf Yes} \\
\bottomrule
\end{tabular}
\end{footnotesize}
\vspace{-0.05in}
\caption{Summary of the strengths and weaknesses of the Bank Fail-Over and Modulo Re-Indexing
schemes, along with \titleShort proposed in this paper.}
\label{summary}
\end{table}


\putsec{mechanism}{Remapping via Native Consistent Hashing}
The key challenges in remapping addresses to handle bank shut-down are
(1) ensuring load-balanced address remapping and (2)
achieving efficient dirty data migration.
This section details how our design leverages key ideas from
consistent hashing to handle the address remapping challenge and to
correctly and efficiently migrate dirty data. Note that CRUNCH focuses
on the remapping mechanism and
can work with any cache
sizing policy 
(e.g.,~\cite{bardine-comparch2007,Naveh:2006,yang-hpca2001,zhang-isca2003}).

\putssec{chash-overview}{Consistent Hashing: A Brief Overview}
Consistent hashing maps addresses to banks indirectly by first hashing
both the banks and addresses on to the same unit
circle. Each address maps to the first available bank encountered in a
clockwise walk of the unit circle starting from its own hash-value on
the unit circle.  As shown in \figref{chash-unitcircle-kevin}(a), address X
maps to bank 2 because it is the first bank encountered in a clockwise
walk (see solid-arrow).  If bank~2 were to be powered down, address X
maps to the next bank in clockwise order, which is bank~3 (see dashed
arrow). We define bank~3 to be the {\em fail-over} bank for bank~2 for
address X.  Note that in this example, any address that maps into the {\em region}
of the unit circle between banks 1 and 2 (e.g., X) maps to bank 2, and
subsequently fails over to bank 3 if bank 2 is disabled.  As shown, this
is identical to BFO.
To provide load balancing, consistent hashing
creates multiple \emph{virtual banks} for each physical bank, such that each
bank is actually associated with multiple regions distributed around the
unit circle. For example,
\figref{chash-unitcircle-kevin}(b) shows three copies of each bank. An
address maps to a bank if it maps to any of the virtual copies of the
same bank. Thus, both X and Y map to bank~2.  Because the virtual
bank copies are permuted in pseudorandom order by the hashing
function, there is not one single fail-over bank for all addresses in
bank~2. For example, the fail-over bank for address Y is bank~1 whereas
the fail-over bank for address X is bank~3.

\figputT{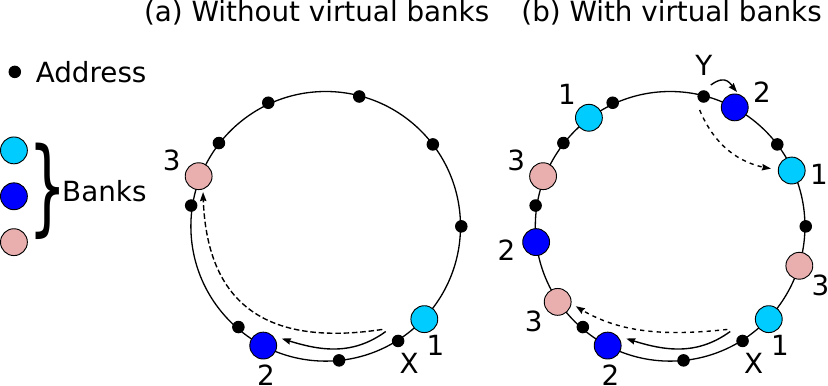}{0in 0in 0in 0in}{Overview of consistent hashing}

Because of symmetry, one can see that the properties are similarly
true for bank power-up. Only a proportional fraction of addresses are
remapped to a newly powered-up bank {\em and} the addresses that are
remapped are distributed among all the previously active banks.

\putssec{multinamespace}{Multi-namespace Variant}

We develop a modified variant of the above consistent hashing
mechanism that facilitates easier hardware implementation in CRUNCH,
while retaining the basic design goals of load-balancing and minimal
remapping. The modified variant introduces two changes in how
addresses and resources (i.e., banks) are mapped to the unit circle.
The first change decomposes 
the mapping of the
addresses to the unit circle into a two stage-process: addresses
are first mapped to ``super-regions'' of the unit-circle, and then they
are further mapped to a region within the super-region.
%
%
\begin{figure}[t]
\begin{minipage}{\linewidth}
\footnotesize
\begin{center}
\includegraphics[clip,scale=0.97]{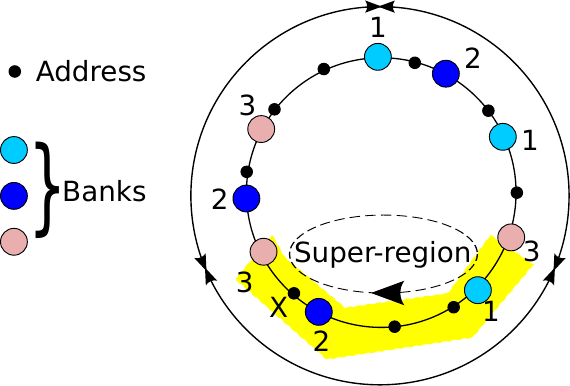}
\end{center}
\caption{Multi-namespace variant of consistent hashing. \label{fig:chash-multi-kevin}}
\end{minipage}
\end{figure}
The second change replaces pseudo-random hashing of resource replicas
with a more-careful construction.  Under the original consistent
hashing, resource distribution may have some non-uniformities. For
example, there may be differences in distribution of the resource
replicas with some parts of the unit circle having higher resource
density than others. Similarly, it is also possible that multiple
replicas of the same resource may be consecutive in clockwise order.
We address this problem by artificially placing exactly one copy of
each resource in each super-region. Consequently, each super-region
has as many regions as resources. The exact order in which the
resources/regions appear in the super-region is pseudo-randomly
permuted.
\figref{chash-multi-kevin} illustrates an example in which the unit circle is
divided into three super-regions. Within each super-region, there is exactly
one virtual instance of each bank.
Further, we modify the fail-over strategy within a
super-region to be limited to resources in that super-region. Thus,
when an address exhausts all resources in a clockwise walk in a
super-region, it wraps around to the first resource in the same
super-region, as illustrated in \figref{chash-multi-kevin}. The address X
originally maps to bank 3. When bank 3 is
unavailable, the address X would map to bank 2 in
traditional consistent hashing because bank 2 appears next in a clockwise
walk of the unit-circle. In our variant, the constraint on
mapping within a super-region forces address X to wrap-around to the
beginning of the super-region and thus maps to bank 1 instead.

\putssec{hardware-remap}{Remapping from a Shut-Down Bank}

The CRUNCH hardware implementation of our consistent hashing variant 
uses a 
{\em region remapping table (RRT)} that contains the mapping from
each region to each physical DRAM cache bank.
We organize the RRT as a two-dimensional
table 
as shown in
\figref{Chash-op}. Along the first dimension, the table holds rows
corresponding to each super-region. Each per-super-region row holds
a wide word that contains a pseudorandom
permutation of all banks. 

The number of super-regions/regions is
a design choice that affects load balance and table size. Because a
region is the smallest unit of the address space
that remaps to other banks, we are
able to achieve better load balance with a larger number of smaller
regions. However this may lead to a larger table.  On the other hand,
with a small number of regions, the RRT can be small and fast; but the
remapping of regions to other banks occurs in chunks that may be too
large to distribute evenly across all banks.  Fortunately, this
tension between remap granularity and table size is not a serious
problem. Because the number of banks is typically small (say, 8 banks
per channel), using 256 total regions (which corresponds to 32~super-regions) is
sufficient to ensure good load balance. Our sensitivity studies with a larger
number of regions (2048 regions in 256 super-regions) show that there is no
significant benefit from larger RRTs.

Our design choices imply a 32-entry RRT where each entry contains 24 bits (8
bank-ids, each of which is 3-bits wide) for a total of 96 bytes, which is small
compared to the DRAM cache. Furthermore, the RRT table does not change the DRAM
internals and it is implemented in the DRAM cache controller.  Note, the RRT is a
read-only table which contains pseudorandom permutations of the bank-ids.
Because it is read-only, it may be computed statically using arbitrary
algorithms at design time.  Furthermore, the problem sizes are fairly small for
offline analysis; we have to choose 32 permutations that are not rotation
equivalent (for our 32-entry RRT) out of 5040 ($= 8!/8 = 7!$) possible
permutations.  Ideally, the permutations must be chosen carefully such that the
fail-over banks are balanced for any number and combination of banks being
shut-down. However, we use a simpler method in which we only ensure that the
fail-over banks are evenly distributed with one bank failure.  Our results show
that our simple table generation method achieves performance very similar to
MRI, which perfectly load balances the different sets across banks, and so we do
not believe that there is significant value in further optimizing the fail-over
orders.

\figputT{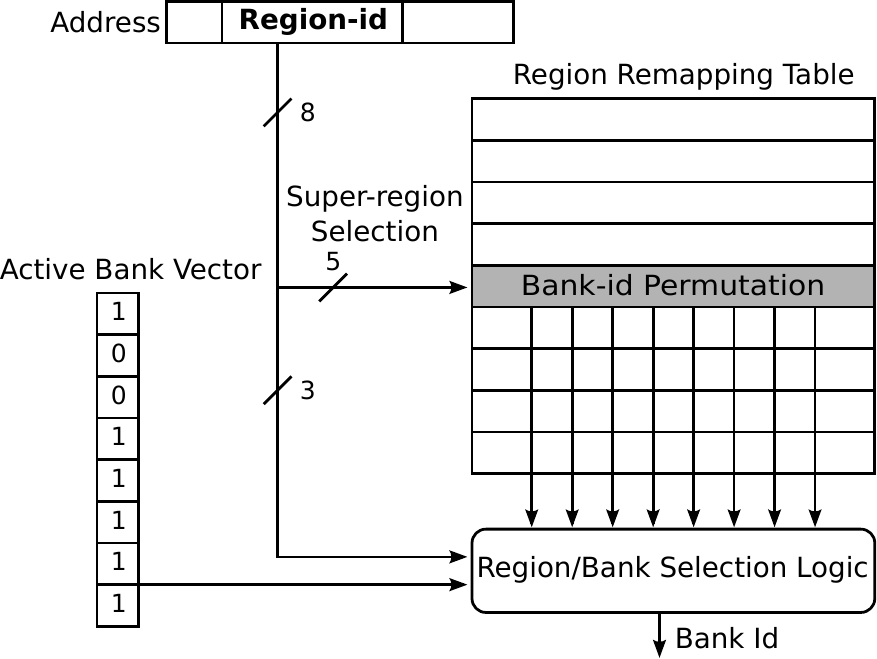}{0in 0in 0in 0in}{CRUNCH hardware}

Each access to the DRAM cache first maps the address to the
appropriate bank by consulting the RRT.  When all banks are active,
the appropriate bank is selected from the permutation by first
indexing into the RRT to read the appropriate super-region entry; and
then selecting the bank by using the additional bits of the
region-index.  When some banks are inactive, we do a
priority-selection to find the first active bank in permutation order
within the super-region.  Because RRT lookup is at hardware speeds,
DRAM cache access latency remains comparable to traditional
bank-selection latency, which typically uses simpler forms of hashing
ranging from trivial bit-selection to bit ``swizzling'' techniques.

\putssec{shutdown}{Handling Dirty Data in a Shut-Down Bank}
When shutting down a bank, dirty data in the bank
must be handled correctly. One simple approach 
writes back all dirty lines to main memory.
We employ an alternative approach that migrates the data from its
home bank (i.e., the bank under the old mapping)
to its new bank. Such migration retains some data in the DRAM
cache while ensuring that dirty data is not ``lost'' to subsequent
reads \footnote{Strictly speaking, correctness is not affected if
lost data is (transitively) dynamically dead. However, we adopt a
strict definition of correctness in which any lost data is treated
as a violation of correctness.}.  The motivation for migrating (as
opposed to writing back) is that write-back operations are limited
by the off-chip memory bandwidth, while migration between banks can
use the DRAM cache's larger bandwidth. Note that there might be a case where
migrating some clean data (e.g., top-N MRU lines) is beneficial for the
steady-state performance (despite an increase in the transition latency). We
leave this as part of future work.



\subsection{Finding the Dirty Blocks}
\label{sec:dirtylocate}
The need to writeback/migrate all dirty data in a DRAM cache bank requires the enumeration of all dirty blocks.
While the dirty bit in traditional write-back cache designs can identify a
given
block as dirty/clean, it cannot offer an efficient {\em enumeration} of
dirty blocks for writeback or migration.
A naive implementation, based on a full walk of the cache bank  to identify and enumerate
dirty data, incurs cost proportional to the size of cache banks even in cases
where the amount of dirty data is a small subset of the data in the cache bank.

We develop an improved implementation with hierarchical dirty bits (HIER)
in which a tree-based hierarchy of dirty row counters indicates dirty-block
presence at successively coarser-grain collections of DRAM rows.  For example,
a root-level dirty block counter indicates the number of rows in the entire
DRAM cache that have at least one dirty block. At the next level of the tree,
a set of $d$ dirty row counters indicate the number of rows in
each of the $\frac{1}{d}$ fractions of the DRAM cache that hold dirty blocks. The
hierarchy ultimately terminates at the individual DRAM cache rows.  Such a
hierarchy enables pruning of the cache walk by avoiding coarse-grain regions of
cache that hold clean data.

HIER incurs two overheads; one in space and the other in time.  The space
overhead is modest because we only count rows that hold dirty blocks as opposed
to the dirty blocks themselves. HIER uses a perfectly-balanced $d$-ary tree that is easy to
represent in linear arrays (similar to $d$-ary heaps).  The number of entries
needed to handle $R$ rows is approximately $(d\cdot R)/(d-1)$ counters.  For
example, consider an $R$=2,048-row DRAM cache bank, with $d=16$.  The root
counter must be 12~bits wide to count up to 2,048 dirty rows. The leaf-node
counters, however, are only 1~bit wide to indicate whether there are any dirty
blocks within each respective row.  Counters at intermediate nodes in the tree
are sized appropriately to count the number of rows covered within that
super-region.  This amounts to only 2,772 bits (346.5 bytes) per table per bank.
Assuming eight banks per channel, and four channels per DRAM stack (see
\secref{meth}), this adds up to 10.8KB for a 128MB DRAM cache. Note that HIER is
not solely used for CRUNCH, but it is also used for BFO and MRI, incurring the
same overhead for all mapping schemes. As we will show in the evaluation
section, HIER provides significant benefit to all remapping schemes.

The HIER structure is implemented as a small SRAM table alongside the
DRAM cache controller and remapping logic.  When a row's dirty counter changes
from dirty to clean (or vice versa), the changes need to be propagated up the
tree to the root. This propagation delay is a negligible time overhead because
(1) $d$-ary trees are fairly shallow, (2) access times to small SRAM arrays are
short compared to DRAM cache access times, and (3) the propagation is not on
the critical path for the main DRAM cache accesses.

\putssec{repatriate}{Repatriating Displaced Dirty Blocks}
\label{sec:repatriate}

A powered-down bank's dirty data may be displaced to several
(potentially all) active banks.  When that bank is eventually
powered up, all its displaced data has to be brought back to the
newly powered-up bank. This need for dirty-data repatriation
imposes two costs; discovery costs to locate such remote dirty
data, and migration costs to copy the data over to the newly
powered-up bank.

To discover dirty data with CRUNCH and MRI, a naive approach is to
walk all banks to discover remote dirty blocks 
(with HIER optimizations to expedite the cache walks).
An alternative 
is to
disallow remote blocks from being dirty (i.e., 
write-through for these blocks only).  However, this may result in steady-state costs in
the form of higher write-traffic to main memory because of
write-through blocks. Because this design increases steady-state
cost to reduce the (uncommon) transient costs, we do not consider
this design in this paper. 
On the other hand, BFO has
an advantage because all displaced dirty blocks that need to be
repatriated can be found in exactly one fail-over bank.

\section{Experimental Methodology}
\label{sec:meth}

\begin{table}[t]
\vspace{2mm}
\begin{footnotesize}
  \centering
    \begin{tabular}{ll}
        \toprule
Processor           & 4-core, 3.2 GHz, 4-wide issue, 256 ROB\\
        \cmidrule(rl){1-2}
Private L1 cache & 4-way associative, 32 KB \\
        \cmidrule(rl){1-2}
Shared L2 cache & 16-way associative, 4 MB\\
        \cmidrule(rl){1-2}
\multirow{4}{*}{Stacked DRAM cache} & 29-way
associative~\cite{loh-micro2011}, 128 MB, \\
 & 1 GHz (DDR 2 GHz), 128-bit channel,\\
 & channels/ranks/banks = 4/1/8, 2 KB rows, \\
 & tCAS/tRCD/tRP = 8/8/15 \\
        \cmidrule(rl){1-2}
\multirow{3}{*}{Off-chip DRAM} & 8 GB, 800 MHz (DDR3 1.6
GHz~\cite{micronDDR3_4Gb}),\\
& 64-bit channel, channels/ranks/banks = 2/1/8, \\
& 2 KB rows, tCAS/tRCD/tRP = 11/11/11 \\
        \bottomrule
    \end{tabular}
  \caption{Configuration of simulated system.}
  \label{table:sysparams}%
\end{footnotesize}
\end{table}

\noindent\textbf{Simulator Model:}
We use MacSim~\cite{macsim}, a cycle-level x86 CMP simulator for
our evaluations of application-level performance. We model a
quad-core CMP system with private L1 caches, and a shared L2
cache. We use a detailed DRAM timing model for the shared L3
\emph{stacked DRAM} cache and the off-chip main memory (DDR3
SDRAM-1600~\cite{micronDDR3_4Gb}). Unless stated otherwise, our
detailed system configuration is as shown in
Table~\ref{table:sysparams}. The size of our DRAM cache is 128 MB.
As we discussed in Section~\ref{ssec:cache-basics}, we assume
the DRAM cache organization proposed by Loh and Hill~\cite{loh-micro2011}.

\noindent\textbf{Power Model:}
We modified Micron's DRAM power calculator~\cite{micron-tr} to
estimate the DRAM cache power. In contrast to off-chip DDR3,
stacked DRAM has a much wider data width of 128 bits. Hence, only
one stacked DRAM chip is accessed for each read or write
command, as opposed to eight chips in the case of x8 DDR3. 
The stacked DRAM therefore requires less energy
per activation because it does not need to access duplicate sets of
peripheral circuits (e.g., row decoders), although we assume the actual
read/write energy itself remains the same because the same total
number of bits are driven.
The DRAM power calculator's IDD values are adjusted accordingly.
Furthermore, we double
stacked DRAM's refresh rate to factor in the higher operating
temperature~\cite{ghosh-micro2007}.

\noindent\textbf{Workloads:}
We evaluate our system using SPEC
CPU2006~\cite{spec}. We focus on memory-intensive
benchmarks because the DRAM cache has
very little impact on low memory-intensity applications.
To form suitable workloads, we group the benchmarks into
categories based on L2 misses per thousand instructions (MPKI).
Benchmarks with MPKI > 25 are in Category H (high intensity),
while those with MPKI > 15 are in Category M
(medium intensity), and the rest of the benchmarks are in Category
L (low intensity).  We construct 10 multiprogrammed workloads
using these benchmarks, as shown in Table~\ref{table:wklds}, for
our main evaluations.

\begin{table}
\vspace{2mm}
\centering
\begin{footnotesize}
\begin{tabular}{lll}
\toprule
\textbf{Mix} & \textbf{Workloads} & \textbf{Category} \\
        \cmidrule(rl){1-3}
WL-1 & mcf mcf mcf mcf & 4H \\
WL-2 & mcf lbm milc libquantum & 4H\\
WL-3 & libquantum mcf milc leslie3d & 4H  \\
WL-4 & milc leslie3d libquantum milc & 4H \\
WL-5 & libquantum milc astar wrf & 2H+2M\\
WL-6 & milc mcf soplex bwaves & 2H+2M \\
WL-7 & milc leslie3d GemsFDTD astar & 2H+2M \\
WL-8 & libquantum bwaves wrf astar& 1H+3M \\
WL-9 & bwaves wrf soplex GemsFDTD & 4M  \\
WL-10 & gcc gcc gcc gcc & 4L \\
\bottomrule
\end{tabular}
\end{footnotesize}
\vspace{-0.05in}
\caption{Multiprogrammed workloads.}
\label{table:wklds}
\end{table}

\noindent\textbf{Evaluating Different Data Remapping Schemes:}
To evaluate the effects of different data
remapping schemes on power and performance, we would ideally like
to run \emph{seconds}-long simulations where banks are powered up
and down dynamically, with data being remapped on every transition.
However, the cycle-level simulation time would be prohibitive
with such an approach and the benefits/trade-offs of different
remapping schemes can be evaluated without such a full-scale
evaluation. Therefore, we decouple the simulation to
separately evaluate the different aspects of the data remapping schemes.
This methodology is similar to those
in other studies for evaluating scenarios spanning
much longer timescales than typical cycle-level simulation sample sizes~\cite{marty-isca2007}.

\ignore{
Ideally, we would like to run \emph{seconds}-long simulations to monitor
applications' working set sizes during the runtime and dynamically power
down/up banks as the working set sizes change throughout the simulation.
However, the simulation time would become prohibitively long with such an
approach. As a result, we decided to decouple the simulation and look at
different aspects of our dynamic power management framework separately.
}

First, we evaluate system performance of different remapping schemes
when different numbers of banks are turned on. This allows us to
examine the performance impact of each remapping scheme's
cache line distribution (load balancing) with a specific bank configuration, for a
certain length of time. We call this the \emph{steady state}
analysis. For evaluating each scheme, we use
a specific number of shut-down banks and simulate 100
million instructions per benchmark. We gather statistics for
each benchmark once it retires 100 million
instructions.  Simulation of all benchmarks continues (to continue to
contend for the memory system) until each one has
executed 100 million instructions.

Second, we evaluate different remapping schemes based on
the transition latency and energy overheads, which we
call the \emph{transition} analysis. We carry out this
analysis by first warming up the DRAM cache with all banks enabled
(for the power down study) or a specified number of banks disabled
(for the power up study). Once the warm-up completes, we power
down (or up) banks, and measure the
latency of the transition and the energy expended to remap dirty
blocks. We run each simulation
for 150 million warm-up cycles, and then execute until the
transition completes.

Third, we generate a representative set of patterns for shutting
down banks instead of using all possible combinations. A policy
that determines how many and which banks to \mbox{power-up/down}
is beyond the scope of this paper as we focus on
the remapping problem. Table~\ref{table:offbank-pattern} shows
the patterns for our evaluations. For each pattern, bank index
starts from left (bank 0) to right (bank 7). The number at each index indicates if
the bank is turned on (\texttt{1}) or off (\texttt{0}).
We use a ``binary search'' approach for selecting bank shut-down order
to avoid (as much as possible) pathological cases where a large number of shut-down banks
all fail-over to the same bank in BFO, thereby creating an unfairly severe hotspot.
We later show
in \secref{naive-pattern} that CRUNCH is
robust against different bank shut-down patterns.
\begin{table}
\centering
\begin{footnotesize}
\begin{tabular}{cc}
\toprule
\textbf{Number of Shut-down Banks} & \textbf{Shut-down Pattern} \\
\cmidrule(rl){1-2}
1 & 11110111 \\
2 & 11010111 \\
3 & 11010101 \\
4 & 10010101 \\
5 & 10010001 \\
6 & 10000001 \\
7 & 10000000 \\
\bottomrule
\end{tabular}
\end{footnotesize}
\vspace{-0.05in}
\caption{Bank Shut-down Patterns.}
\label{table:offbank-pattern}
\end{table}

\noindent\textbf{Performance Metrics:} For the steady-state
analysis, we report system performance using the \textit{Weighted
Speedup} metric:
{\centering $Weighted Speedup = \sum_{i=1}^N{\frac{IPC_{i}^{shared}}{IPC_{i}^{alone}}}$~\cite{harmonic_speedup,weighted_speedup}}

\noindent For the transition analysis, we report the latency of
powering up and down banks in \emph{cycles}. We also
report DRAM energy consumption during power-up/down
transitions.

\putsec{eval}{Experimental Evaluation}


\subsection{Steady-State System Performance}
\label{sec:steady-state}

In this section, we evaluate the impact of different
remapping schemes on system performance in terms \emph{weighted speedup} during
steady state operation. Specifically, we shut down different
numbers of banks and examine the load-balancing effect of each
remapping scheme.
\figref{isca/steady-ws-crop.pdf} shows the steady-state system
performance of each remapping scheme averaged (geometric mean)
across all workloads for different numbers of active banks. We
draw two key conclusions.

First, CRUNCH, with carefully designed region remapping tables,
provides performance comparable to MRI. Therefore, we conclude
that CRUNCH achieves good load balancing, on par with MRI. In
fact, except for the case of three enabled banks, CRUNCH provides slightly
better average system performance than MRI. While MRI provides
even distribution over the address space, the simple
\emph{modulo-k} mapping means that accesses at strides that are
not relatively prime to the number of enabled banks will map to a
restricted subset of the available banks. In contrast, the RRT
lookup in CRUNCH performs a swizzling that will evenly distribute
these strided accesses across all available banks. As a result,
we expect that MRI coupled with a reasonable swizzling scheme
would perform on par with CRUNCH in the steady state.

Second, CRUNCH outperforms BFO when at least one
bank is powered down. This is because BFO creates unbalanced load
distribution by remapping all cache lines within a powered-down
bank to the next available consecutive bank. For instance, when
one bank is powered down, the consecutive bank will receive
roughly twice as much request traffic as the other remaining
banks. To quantitatively compare the load balancing effect between
different remapping schemes, we define a metric called the
\emph{imbalance ratio}. The imbalance ratio is the maximum request
traffic to any bank divided by the minimum request traffic to any
other bank.  With three active banks, we observe the imbalance
ratios are 4.2, 1.3, and 1.0 for BFO, CRUNCH, and MRI,
respectively, on average across all workloads. The high bank
imbalance leads BFO to reduce average system performance by 11\%
compared to CRUNCH.
In summary, CRUNCH provides the ability to consistently
provide low imbalance ratios, enabling high system performance,
for all workloads. Thus, we conclude that CRUNCH achieves the
design goal of effectively load-balancing cache line distribution
across banks.

\figputT{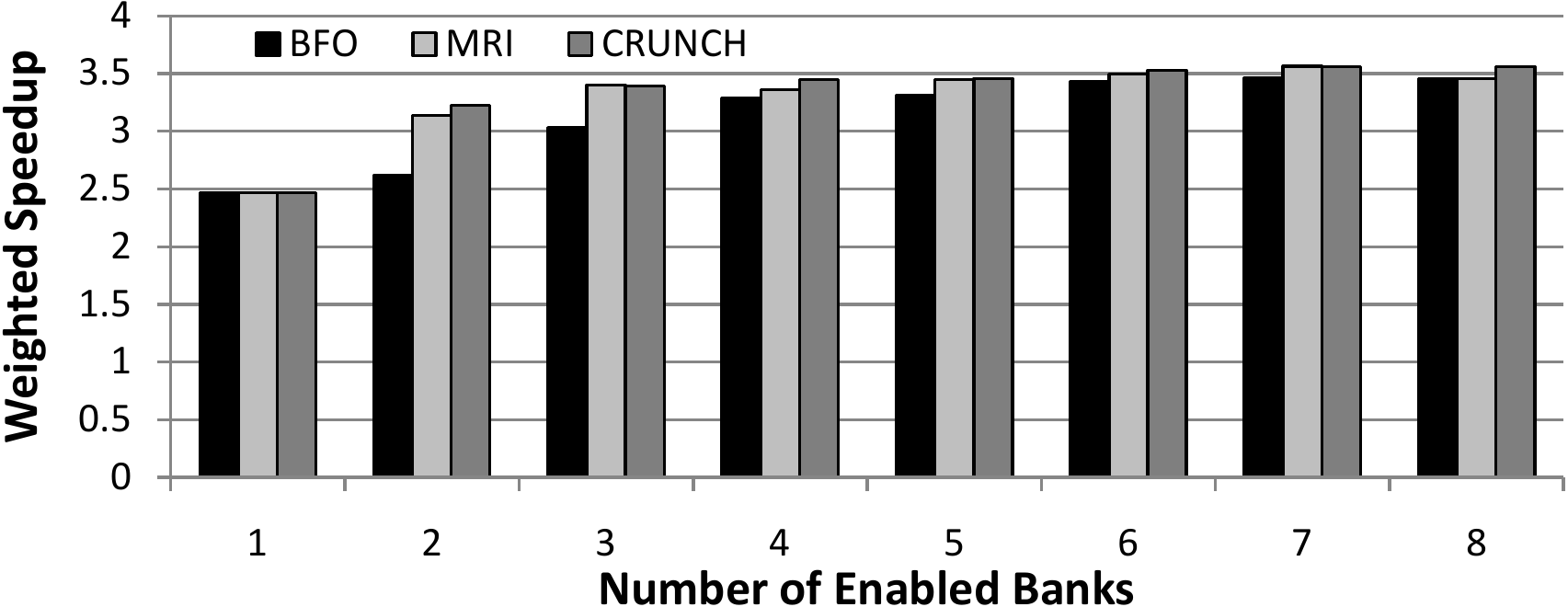}{0in 0in 0in 0in}{Steady-state system performance
(weighted speedup) of the evaluated bank remapping schemes across varying
numbers of enabled DRAM cache banks.}

\figputT{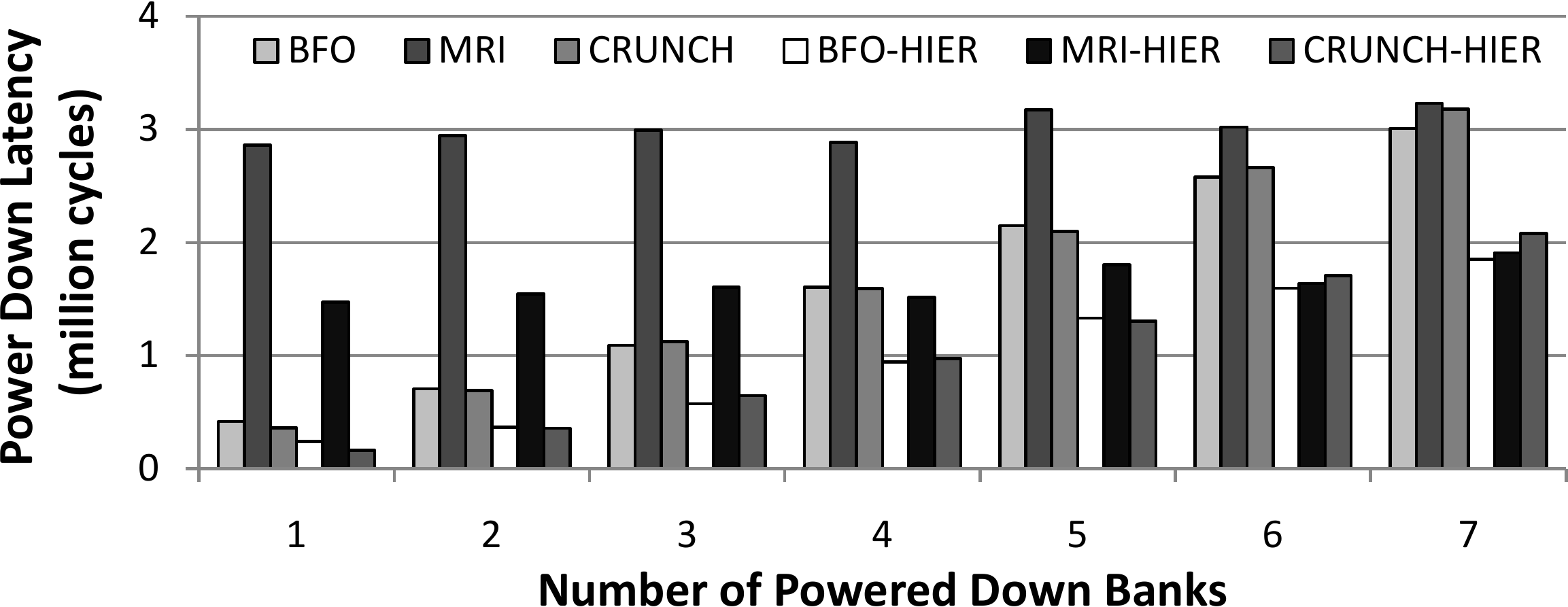}{0in 0in 0in 0in}{The power-down transition
latencies of remapping schemes for varied numbers of powered-down
banks.}


\subsection{Power-Down Transition Analysis}

\noindent\textbf{Latency Analysis:}
There are two phases in powering down a cache bank. The first
phase is searching the bank for all modified cache lines so that they
can either be migrated to another bank or written back to main
memory.  The second phase is actually performing the data
transfer. The latency associated with performing these operations
is a factor in system performance as well as in determining how
responsive a power management scheme can be to short-term changes
in workload behavior.

\figref{isca/down-all-crop.pdf} shows the power-down transition
latency averaged across all evaluated workloads for shutting down
different numbers of banks (starting with all eight banks enabled), with and without the proposed
hierarchical dirty bits technique (\hier). Two conclusions are in
order. First, CRUNCH achieves significantly lower power-down
latencies than MRI, on par with BFO. Second, as the number of
powered-down banks decreases, the transition latency also drops
for CRUNCH and BFO, whereas the transition latency remains
approximately the same for MRI. This is because CRUNCH and BFO
only need to walk through and transfer dirty cache lines in the
powered-down banks. In contrast, MRI requires walking through all
banks because \emph{all} cache lines are remapped across the
remaining active banks after banks are powered down.

\noindent\textbf{Effect of hierarchical dirty bits:}
\figref{isca/down-all-crop.pdf} shows that using the proposed
hierarchical dirty bits technique (\hier) provides consistent
latency reductions for all workloads and variations of
powered-down bank counts, with maximum reductions of 49\%, 55\%,
and 49\% for BFO, CRUNCH, and MRI, respectively. This is because
using \hier avoids unnecessary cache walks by enumerating dirty
lines directly for migrations.

\noindent\textbf{Transition impact on instantaneous performance:}
To gain a better sense of how the power-down transition impacts
instantaneous performance and system responsiveness,
\figref{isca/down-short-crop.pdf} and
\figref{isca/down-long-crop.pdf} show detailed performance traces
sampled over time for two representative workloads, \mbox{WL-1}
and \mbox{WL-2}, when shutting down two banks. The
y-axis indicates the total number of retired instructions during the
last sampling period (500K cycles). Several conclusions can be
drawn from this data.

First, the system performance drops significantly regardless of
the remapping schemes used for both workloads. The reason is that
the DRAM cache is prevented from servicing memory requests during the transition
to guarantee correctness.

Second, MRI has a much larger transition latency than both BFO and
CRUNCH. There are two main reasons for this. First, MRI requires
walking through the whole cache to enumerate dirty cache lines
because cache lines are uniformly distributed across all active
banks. Second, MRI also needs to migrate more dirty cache lines
than BFO and CRUNCH because of the same explanation as the first
reason: uniform cache line redistribution.
For instance, the
number of migrated dirty cache lines are 817, 842, and 2606, for
BFO, CRUNCH, and MRI, respectively, for WL-1's example transition in \figref{isca/down-short-crop.pdf}. As a result, the
migration latency is significantly higher for MRI.
In addition,
MRI provides lower system performance during the initial program
phase right after the completion of power-down transitions for
both \mbox{WL-1} and \mbox{WL-2}. For example,
\figref{isca/down-short-crop.pdf} clearly shows that the retired
instruction curve for MRI falls below BFO and CRUNCH after the
transition completes at the sample point 305. This occurs because
MRI remaps a large number of clean cache lines. Therefore, most of the requests
will result in misses during the program phase after the completion of
power-down transitions.

Third, the transition overhead for \mbox{WL-2} is greater than
that for \mbox{WL-1} simply because \mbox{WL-2} has a much larger
quantity of dirty data, thus requiring longer latency for
transferring these dirty cache lines.
For WL-2, the number of
migrated dirty cache lines are 134937, 120630, and 393366, for
BFO, CRUNCH, and MRI, respectively, which is at least 100x more
than WL-1.



\figputT{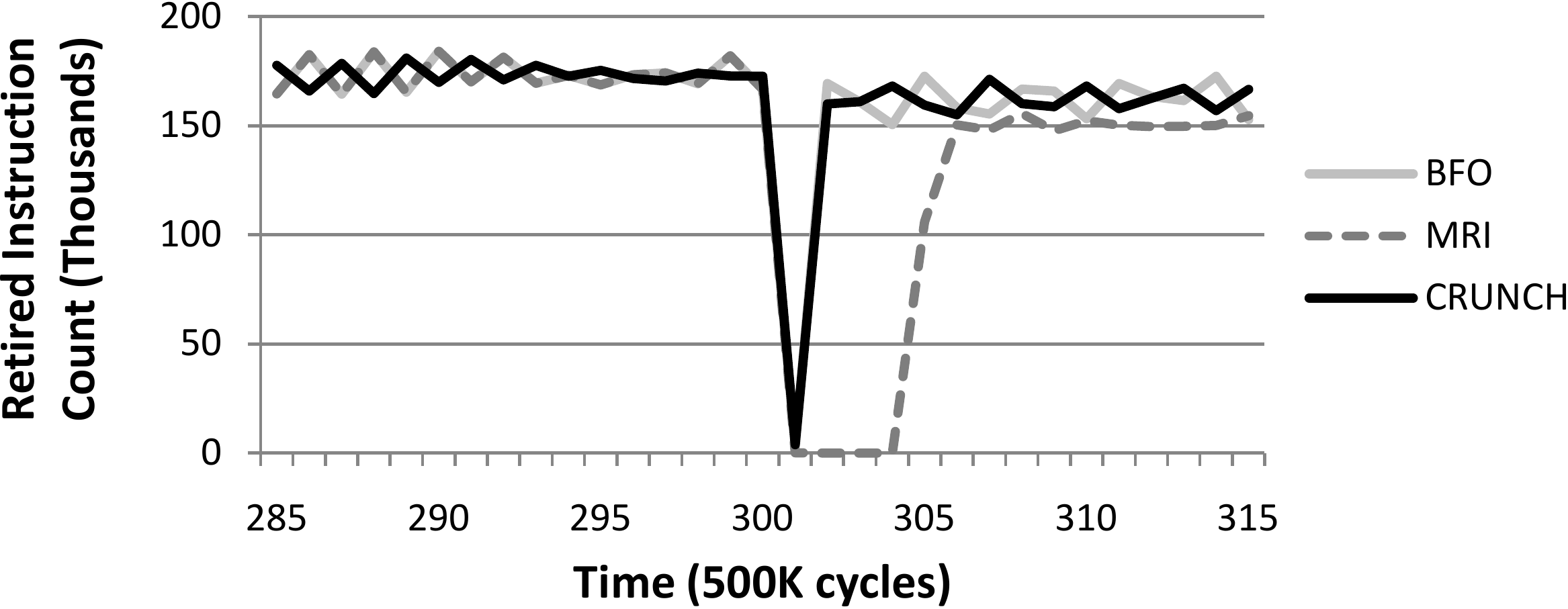}{0in 0in 0in 0in}{Performance of WL-1
sampled over time.}
\figputT{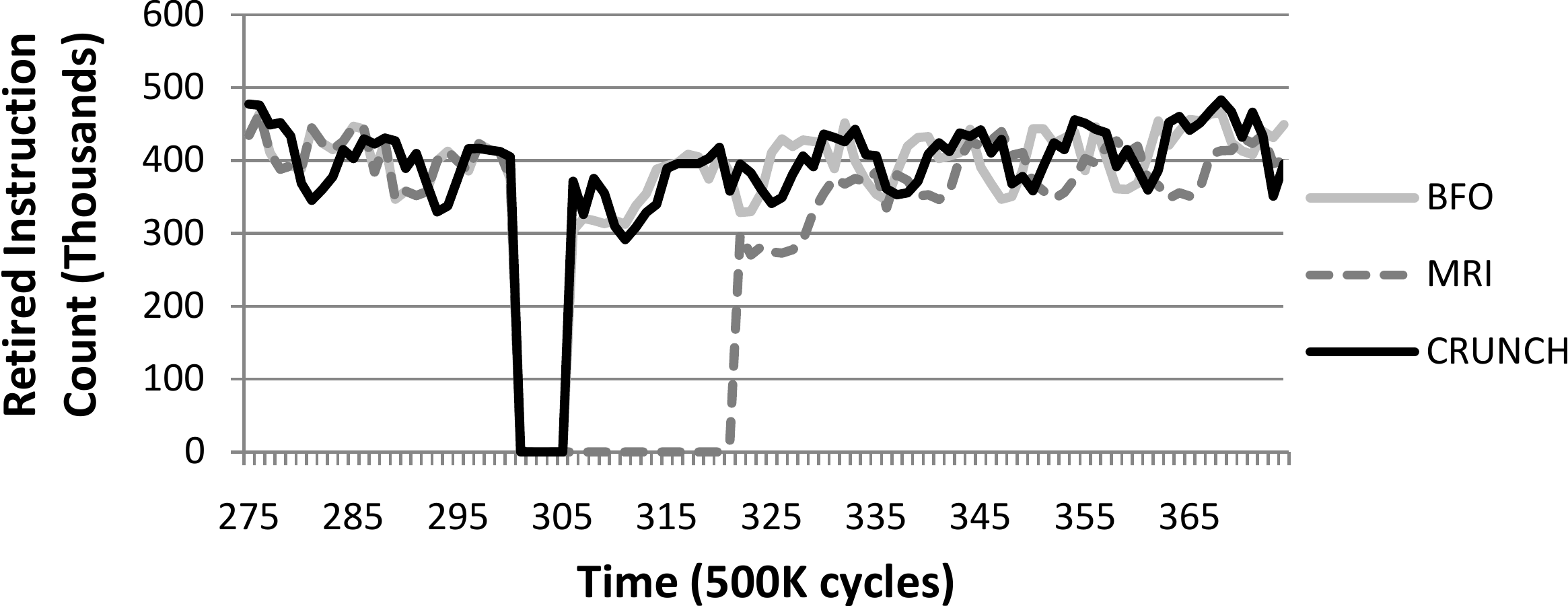}{0in 0in 0in 0in}{Performance of WL-2
sampled over time.}

\noindent\textbf{Energy Analysis:}
\figref{isca/down-energy-crop.pdf} shows the energy consumption
for each remapping scheme with HIER applied as the number of
powered-down banks is varied.  Because energy consumption is
proportional to the number of dirty cache line
migrations, the energy numbers follow the same trend as the
latency numbers.  The following are some major conclusions. First,
BFO and \titleShort consume significantly lower energy for lower
numbers of powered-down banks, as they need to migrate only the
dirty cache lines in the powered-down banks. On the other hand,
MRI's power consumption remains roughly the same for all bank
configurations because MRI remaps data in \emph{all} banks.
Second, \titleShort actually consumes 3.8\% and 8.8\% more energy
than MRI when shutting down six and seven banks, respectively.
This is because \titleShort needs to migrate 3.6\% and 7.6\% more
dirty lines when shutting down six and seven banks respectively,
than MRI for these two configurations.  Nonetheless, \titleShort
consumes 8.6x less energy than MRI in the best case (shutting down
one bank).

\figputTT{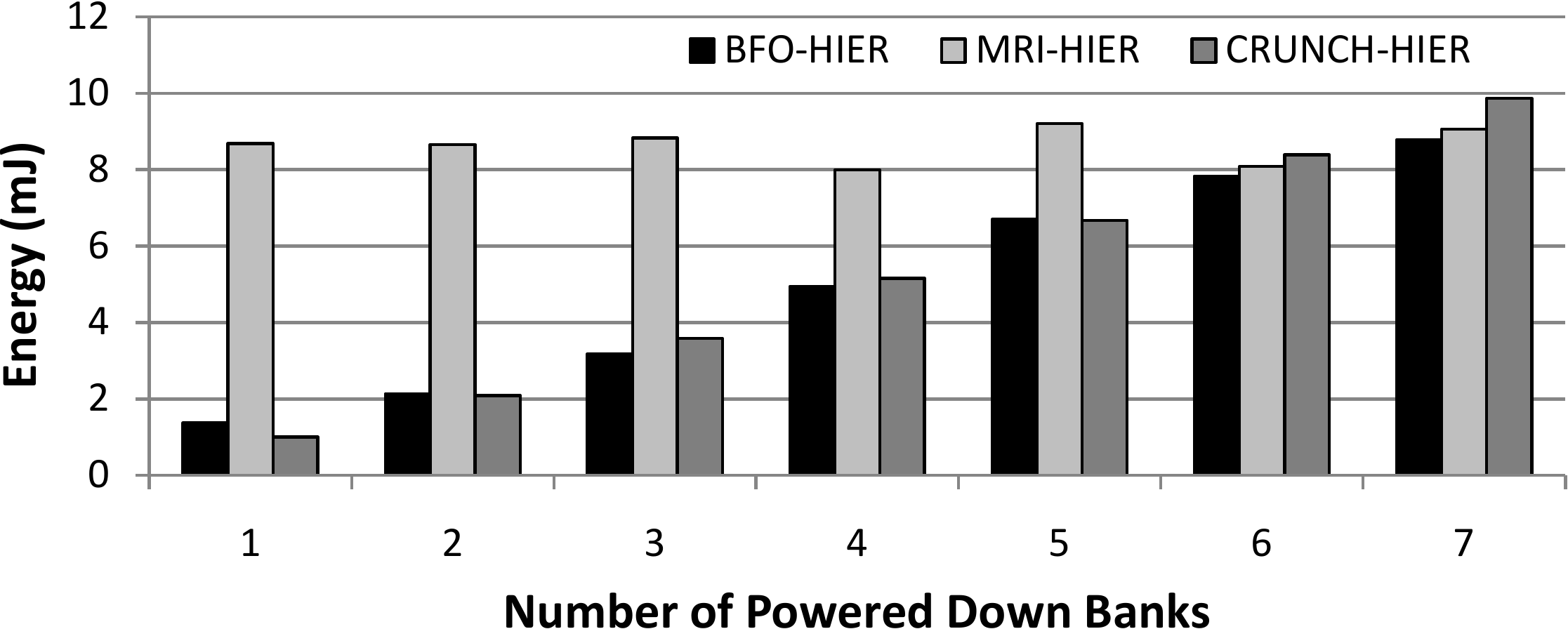}{0in 0in 0in 0in}{Energy consumption during
power-down transitions}

\subsection{Power-Up Transition Analysis}

\noindent\textbf{Latency Analysis:}
Powering a cache bank back up requires finding all of the modified
cache lines that previously would have mapped to this bank, but
were displaced elsewhere. This requires searching one or more
banks, and then either repatriating those cache lines back to this
bank, or writing them back to main memory.
\figref{isca/up-all-crop.pdf} shows the power-up transition
latency (when powering back up to all eight banks enabled) averaged across all
evaluated workloads for different remapping schemes along with the \hier
mechanism. We draw the following conclusions.

First, using BFO consistently provides lower transition latency
for powering up banks, compared to both CRUNCH and MRI across all
variations of powered-up bank counts. As explained in
\secref{repatriate}, this is because BFO only needs to search
through a subset of active banks that contain failed-over cache lines.
Finding these active banks is straightforward because they are the
next sequential active bank(s) relative to the newly powered-up
bank(s). On the other hand, both CRUNCH and MRI require searching
every active bank to find the displaced dirty cache lines. As a
result, when a naive cache line walking scheme that reads
all cache lines in all active banks is employed, the latency of
reading every cache line in the active banks to find displaced
dirty bits dominates the transition latency for CRUNCH and MRI.

Second, similar to our observations for the power-down transition
analysis, using a smarter cache line walking scheme
(\hier) reduces the transition latency
significantly for all remapping schemes.  The maximum reductions are
66.2\%, 69.3\%, and 66.2\% for BFO, CRUNCH, and MRI, respectively.


\noindent\textbf{Energy Analysis:}
\figref{isca/up-energy-crop.pdf} shows the energy consumption
for each remapping scheme with \hier applied, as the number of
powered-up banks is varied. Following are our key conclusions.
First, the energy consumption is proportional to the transition
latency as explained before, for powering down banks. Second,
\titleShort consumes less energy than MRI for all bank
configurations since MRI requires remapping all data when the bank
configuration changes. Third, BFO consistently provides lower
energy consumption than both \titleShort and MRI. This
behavior is the same as for transition latency, which we explain
in detail above. Nevertheless, \titleShort provides
better system performance than BFO during steady-state.

Although BFO provides the lowest power-up transition latency and hence transition energy among
all the remapping schemes, as demonstrated in
Section~\ref{sec:steady-state}, it has the disadvantage of leading
to potentially unbalanced cache line distribution, causing some
banks to become hotspots and hence become bottlenecks for
memory requests to these banks. Therefore, BFO's lower power-up
latency comes at the cost of reduced steady-state system
performance. In contrast, CRUNCH not only provides high system
performance with load-balanced distribution of cache lines, but
it also enables fast transition latency for powering down banks and
offers lower power-up transition latency than MRI. By
applying a simple optimization, hierarchical dirty bits (\hier),
the power-up transition latency gap between BFO and CRUNCH
significantly reduces. In addition, other optimizations that can
more efficiently enumerate dirty data or bound the amount of dirty
data in the cache, could potentially further reduce the transition latency
gap between BFO and CRUNCH.
Therefore, we conclude that CRUNCH achieves the two design goals
of enabling low transition latency and providing load-balanced
cache line distribution.

\figputTT{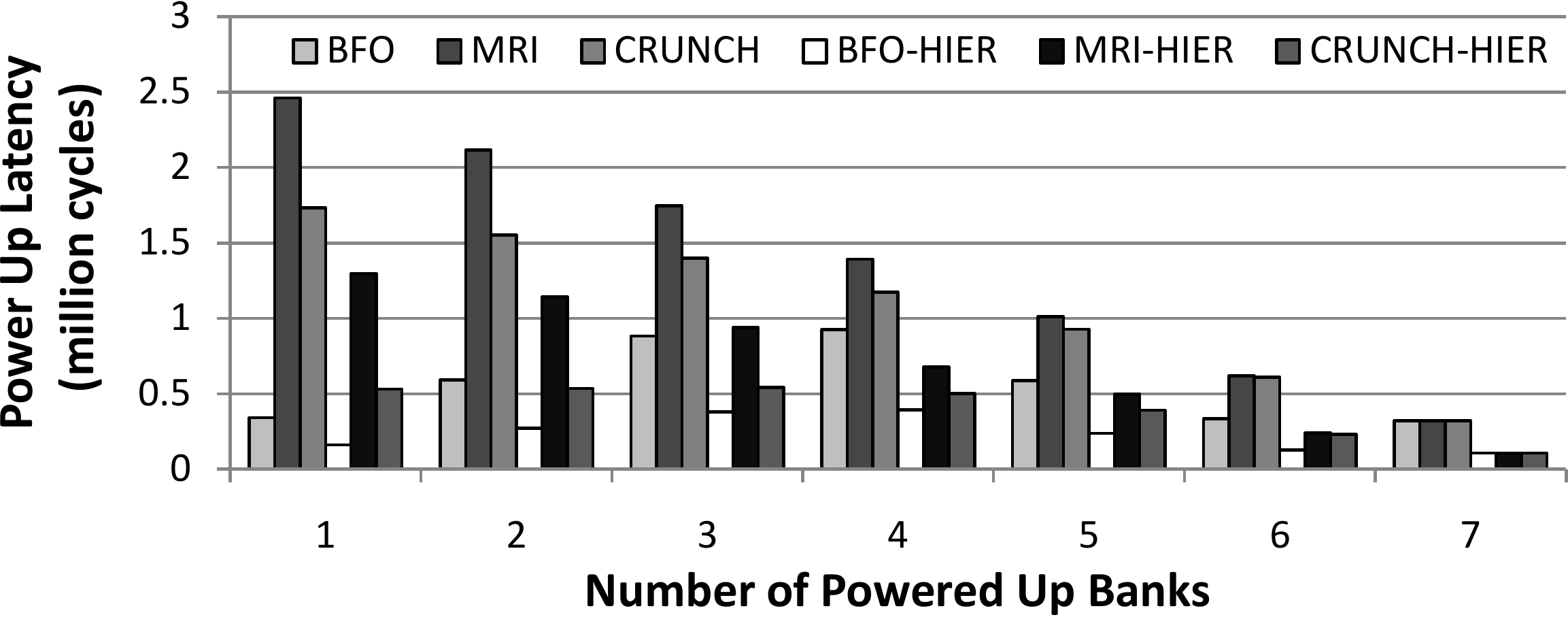}{0in 0in 0in 0in}{The power-up transition
latencies of different remapping schemes averaged across all workloads
for turning on different numbers of banks.}

\figputTT{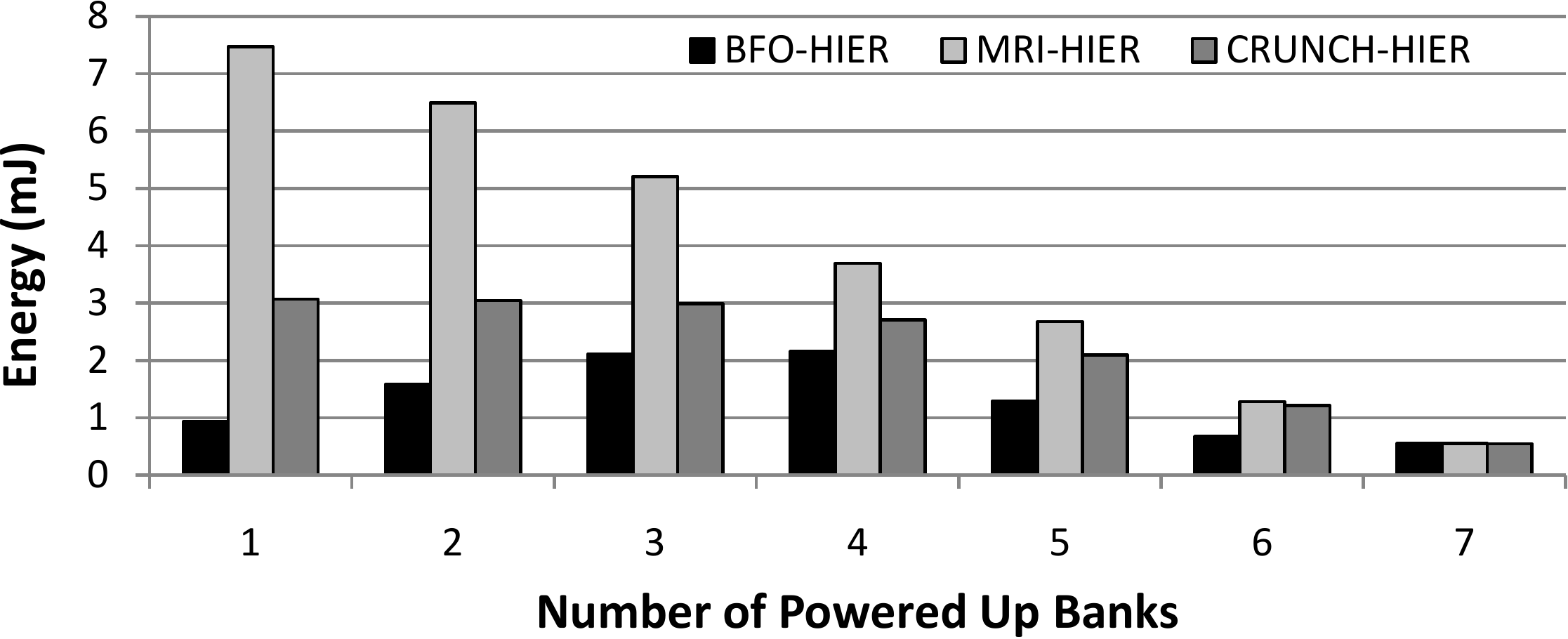}{0in 0in 0in 0in}{Energy consumption during
power-up transitions}

\subsection{Sensitivity to System Parameters}
\label{sec:sense-cache}

\ignore{
\noindent\textbf{Varying Cache Size:}
\kevin{The ws shows that there is very little performance impact. It seems like
128MB is enough to fit the working sets. Do we still want to briefly talk about
it?}
}

\noindent\textbf{Varying Core Count:}
\figref{isca/steady-ws-8core-crop.pdf} shows the average
\mbox{steady-state} system performance of different remapping schemes on an
8-core system running eight randomly-picked multiprogrammed
workloads. Similar to the observation we made for our baseline
4-core system in Section~\ref{sec:steady-state}, \titleShort
provides slightly better system performance than MRI and continues
to outperform BFO. Therefore, we conclude that \titleShort
continues to provide both good steady-state performance and low
transition latency/energy with increasing cache pressure.

\figput{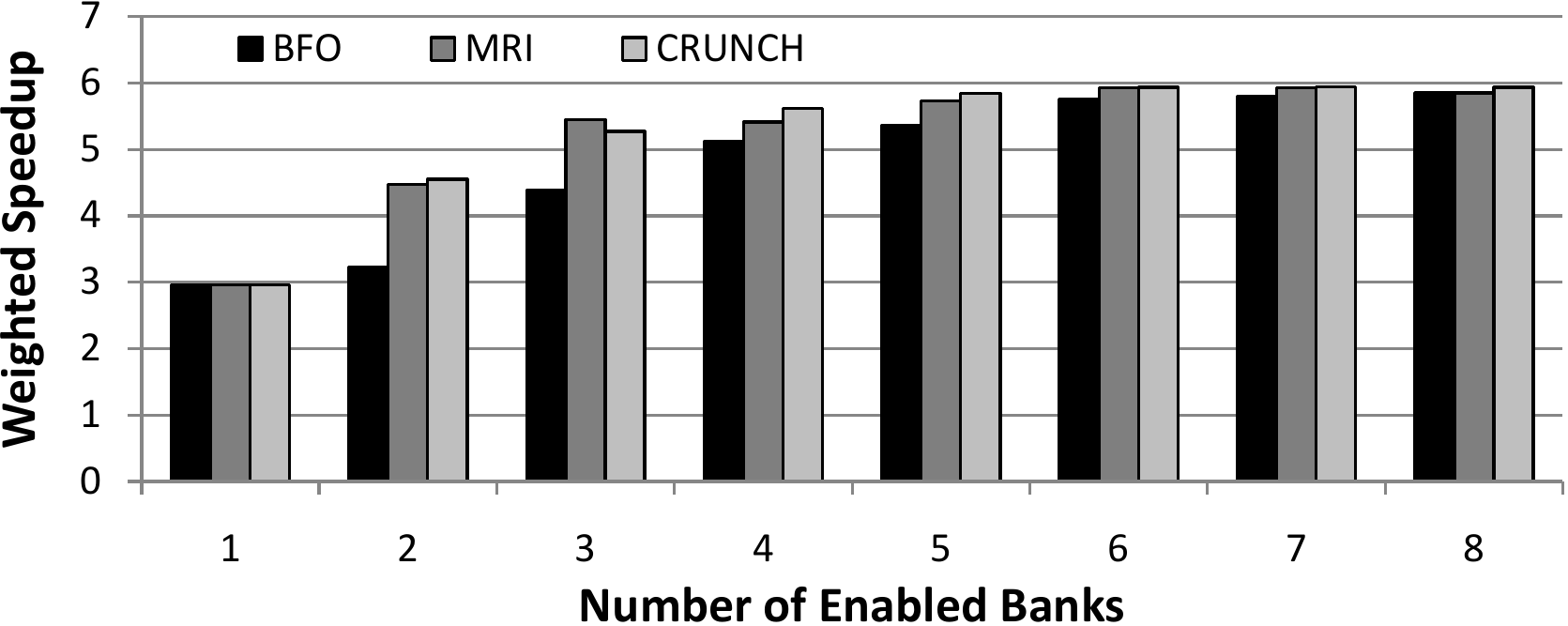}{0in 0in 0in 0in}{System performance of
different remapping schemes on an 8-core system.}

\subsection{Impact of Sequential Bank Shut Down}
\label{sec:naive-pattern}
In this section, we evaluate the impact of different bank shut-down
patterns on steady-state system performance. For our evaluations
so far, we have been using balanced bank shut-down patterns,
described in Section~\ref{sec:meth}.
Alternatively, we can power down banks sequentially
starting from the bank with the lowest index value.
\figref{isca/steady-seq-ws-crop.pdf} shows the steady-state
performance comparisons between sequential and balanced bank
shut-down orders.
We make the
following observations. First, BFO significantly reduces system
performance when banks are being turned off sequentially. This is
because BFO creates an unbalanced load distribution by remapping all
the cache lines within the shut-down banks to a single active bank.
Second, CRUNCH provides the same system performance compared to
that of using balanced shut-down orders even if banks are
sequentially disabled. The importance of these results is that with
\titleShort, the overall DRAM cache power management policy can be
simplified so that it needs only figure out the {\em number} of
banks that should be turned on, rather than which {\em specific} banks.

\figput{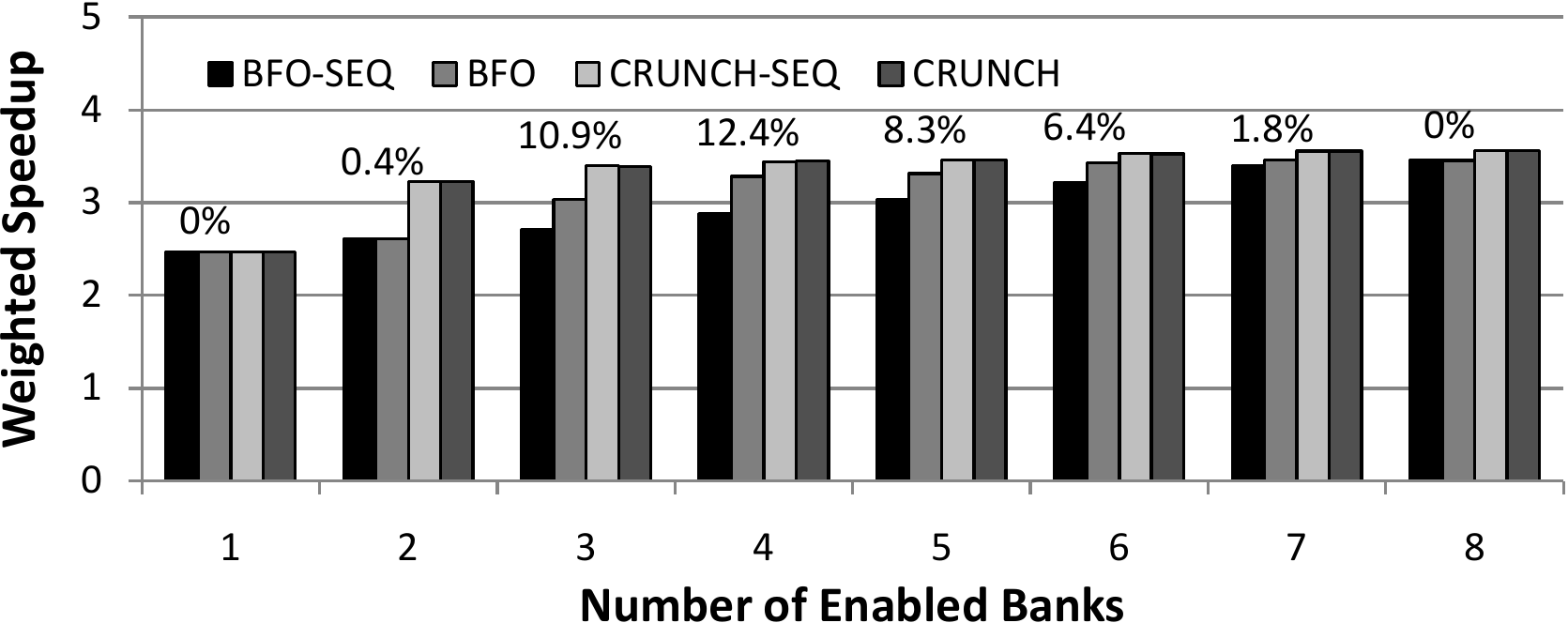}{0in 0in 0in 0in}{Steady-state performance comparisons
between sequential- and balanced-order bank shut-down for BFO and CRUNCH. \%
values are performance degradations of BFO-SEQ relative to BFO.}



\section{Related Work}
\label{sec:related}

While techniques to turn off banks have been studied by several
works in the context of SRAM caches, the organization of DRAM
caches poses additional challenges. Specifically, entire sets are
mapped to a row in a bank. Therefore, powering down banks requires
(1) remapping of the sets mapped to the being powered-down banks,
to the active banks and (2) migration of dirty data from the being
powered-down banks to the active banks. Naive schemes to perform
this remapping/migration suffer from the problem of either high
transition times (to remap/migrate) or degrade performance in the
steady state (after remapping/migration).

To our knowledge, this is the first work to propose a data remapping
scheme that achieves both good load-balancing in the steady state
and low bank power-up/down transition latency/energy.

A number of prior works have proposed to save SRAM cache power by
disabling cache ways (e.g.,~\cite{albonesi-micro99,
powell-islped2000, zhang-isca2003}). Specifically,
Albonesi~\cite{albonesi-micro99} proposes to selectively disable
cache ways in the L1 data cache to reduce dynamic power
consumption. Bardine et al.~\cite{bardine-comparch2007} propose a
D-NUCA cache that dynamically turns on/off ways based on the
running application's behavior, to save both dynamic and static
power. Zhang et al.~\cite{zhang-isca2003} propose a configurable
cache design that allows varying the associativity to save dynamic
power.

Industry processors have also adapted dynamic cache sizing
techniques to reduce leakage power. The Intel Core Duo
Processor~\cite{Naveh:2006} implements a dynamically sizeable L2
cache which is down when the processor core enters deep sleep
states. Similarly, AMD's Steamroller processor~\cite{SteamRoller}
implements L2 cache resizing based on the workload.
Both of these implementations disable ways of the cache to achieve
a smaller size.

All of these works focus on small SRAM cache designs that have
separate tag stores and whose ways are distributed across all
banks. Therefore, turning off banks merely reduces the
associativity of the cache. Our work, on the other hand, focuses
on a large DRAM cache, for which a practical organization requires
an entire set to be mapped to a row in a bank, as described in
Section~\ref{ssec:cache-basics}. Therefore, when turning off banks
in such DRAM caches, the sets mapped to a bank need to be remapped
to other active banks.

Prior works have proposed to dynamically adapt the cache size at a
finer granularity by disabling cache lines using decay-based
techniques (e.g.,~\cite{abella-taco2005, hanson-vlsi2003,
hanson-iccd2001,hu-tocs2002, kaxiras-isca2001, monchiero-icpp2009,
zhou-pact2001}). These are complementary to our proposed bank
shut-down scheme and can be applied to DRAM caches as well.

Yang et al.~\cite{yang-hpca2002} propose a hybrid technique that
dynamically resizes caches to save leakage power by changing the
number of active cache sets~\cite{powell-vlsi2001, yang-hpca2001}
or ways~\cite{albonesi-micro99}. Following are the key reasons why
CRUNCH is significantly different from this work. First, this
mechanism only allows power-of-two resizing in order to address
the indexing issue when enabling or disabling sets. This scheme is
similar to MRI with a power-of-two resizing and would result in a
large number of misses, since a number of cache blocks would have
to be remapped. Second, the mechanism's primary focus is a
policy to determine the number of sets or ways to dynamically
resize, which is not the focus of our work.

There has also been significant work in reducing power in off-chip
DRAMs
(e.g.,~\cite{Cooper:Micro10,Delaluz:HPCA01,hegde-codes2003,Hur:HPCA08,Udipi:ISCA10}).
These works employ different techniques such as partial activation of the
DRAM array or low energy operating modes to reduce DRAM power and
do not address the challenges in turning off banks when DRAM is
used as a cache.

\section{Conclusion}

Die-stacking technologies enable stacking a processor with DRAM
and this DRAM can act as a large, high-bandwidth cache. While some
workloads utilize the entire capacity of this large DRAM cache,
several workloads use only a subset of this cache. Therefore,
powering down banks of this DRAM cache can enable power savings,
without degrading performance significantly. However, since entire
sets are mapped to the same row (in a bank) in a typical DRAM
cache, when a bank is powered down, the sets mapped to it need to
be remapped to other active banks. Our goal in this work is to
address this data remapping challenge, thereby enabling DRAM cache
resizing. To this end, we presented Cache Resizing Using Native
Consistent Hashing (CRUNCH), a data remapping scheme that achieves
both low transition latency {\em and} load-balanced cache line
distribution. CRUNCH outperforms two state-of-the-art remapping
schemes, providing \emph{both} (1) high system performance during steady-state
operation, with a certain number of banks turned on and (2) fast
transition latency and energy when banks are powered up or down.
Therefore, we conclude that CRUNCH provides an effective data
remapping substrate, enabling DRAM cache resizing.

\ignore{
when DRAM cache banks are dynamically powered down or up. First, CRUNCH provides the
benefit of low transition latency by only migrating dirty data blocks in the
shutdown bank without affecting cache lines in other active banks. Second, to
achieve load-balanced distribution of the migrated cache lines among the
powered-up banks, CRUNCH maps addresses to \emph{regions}, each
with a fail-over order which is a unique permutation of
bank indices.   An address selects a bank from this permutation, and if
an address accesses a bank that is powered down, the address
maps to the next available (powered) bank index in the fail-over order. By
using carefully generated permutations that ensure fail-over bank indices are
evenly distributed with one bank failure, CRUNCH is able to achieve a fair
load-balancing cache line distribution across banks, thus providing high system
performance in the presence of powered-off banks. CRUNCH outperforms
two different mapping schemes, while providing low transition latency when banks are powered down or up.
We conclude that CRUNCH is an effective high-performance mapping scheme for
dynamic cache power management.
}


{
\bstctlcite{bstctl:etal, bstctl:nodash, bstctl:simpurl}
\bibliographystyle{IEEEtranS}
\bibliography{paper}
}

\end{document}